\documentclass[preprint,prd,aps,showpacs,showkeys,nofootinbib]{revtex4}
\usepackage{graphicx}
\begin{document}


\title{The LHC 750 GeV diphoton excess in supersymmetry with gauged baryon and lepton numbers}

\author{Tai-Fu Feng$^{a}$\footnote{email:fengtf@hbu.edu.cn},
Xue-Qian Li$^b$\footnote{email:lixq@nankai.edu.cn},
Hai-Bin Zhang$^{a}$\footnote{email:hbzhang@hbu.edu.cn},
Shu-Min Zhao$^a$\footnote{email:zhaosm@hbu.edu.cn}}

\affiliation{$^a$Department of Physics, Hebei University, Baoding, 071002, China\\
$^b$Department of Physics, Nankai University, Tianjin, 116024, China}

\begin{abstract}
The significance of discovering the boson of 750 GeV is beyond
finding a single heavy boson, because it may hint the location of
the scale for new physics beyond the standard model which is
the target of long-time exploration. There have been many models to
explain the diphoton  excess observed by the ATLAS and CMS
collaborations and the BLMSSM is one of them. The BLMSSM is an
extension of the minimal supersymmetric model where baryon
and lepton numbers are local gauge symmetries. We analyze the decay channels
$\Phi\rightarrow gg$, $\Phi\rightarrow\gamma\gamma$, $\Phi\rightarrow Z\gamma$,
and $\Phi\rightarrow \bar{t}t,\;VV\;(V=Z,\;W)$ with the mass of the
CP-odd scalar $\Phi=A_{_B}^0$ being around $750\;{\rm GeV}$ in this model. Within
a certain  parameter space, the scenario can account for the
experimental data on the diphoton excess.
\end{abstract}


\maketitle

\section{Introduction\label{sec1}}
\indent\indent  A new resonance with its mass around $750\;{\rm
GeV}$ has been observed at the LHC
at a center of mass energy of $13\;{\rm TeV}$ through the process
$pp\rightarrow\Phi\rightarrow\gamma\gamma$~\cite{LHC13,LHC13-1}. If this
observation is confirmed by subsequent experiments, the excess
certainly manifests a signal of new physics beyond the standard model (BSM) and would be a milestone
for high energy physics. Even though the standard model (SM) is very
successful and almost all of its predictions are consistent with the
standing experimental data, it is known that the SM is an effective
theory of some underlying principles. So far nobody knows what the
underlying principle is and a more bothersome situation is that there
was not any hint about where the scale for the new physics should
be. Therefore, besides looking for the SM Higgs which is the base of
our SM, the second target of LHC is to search for new physics.
The first target was fulfilled and the 125 GeV
SM Higgs was discovered, thus the attentions of all physicists are
turned to look for new physics, at least we need to determine
the scale of new physics which would provide valuable information
for building next generation of accelerators.

There have been many models beyond the SM and most of them possess a
scalar or pseudoscalar boson(s) which may stand as the 750 GeV
observed at LHC and  be responsible for the diphoton
excess.
For example, in Refs.~\cite{NSoneS1,NSoneS1-1,NSoneS2,NS1,NSoneS3},
a scalar particle with $m_\Phi=750$ GeV is introduced which may decay into two photons
as $\Phi\rightarrow 2 \gamma$. Alternatively, in the framework of a
minimal UV-complete model with a massive singlet pseudoscalar
state, this diphoton excess is discussed \cite{NS2}. Several models
containing exotic fermions (a single vector-like quark with charge 2/3e, a
doublet of vector-like quarks, a vector-like generation including
leptons) are considered, and theses particles can contribute to the
$\Phi\rightarrow 2\gamma$ \cite{NS3}. With the supposition that
vector-like quarks or leptons strongly couple to
the heavy Higgs and photons or gluons in those new models, the
diphoton resonance at a mass of 750 GeV~\cite{NSVQ} can be
explained. Possible relations between the newly observed resonance
and the dark matter are analyzed in the works of \cite{NSdark,NSdark-1}.
There are also other works
~\cite{NSother-0,NSother,NSother-1,NSother-2,NSother-3,NSother-4,NSother-5,NSother-6,
NSother-7,NSother-8,NSother-9,NSother-10,NSother-11,NSother-12,NSother-13,
NSother-14,NSother-15,NSother-16,NSother-17,NSother-18,NSother-19,NSother-20,
NSother-21,NSother-22,NSother-23,NSother-24,NSother-25,NSother-26,NSother-27}
which also research
the diphoton excess reported  by the the CMS and ATLAS
collaborations. Up to now, there can be found hundreds of papers about it in arXix.

As the simplest soft broken supersymmetry theory, the minimal
supersymmetric extension of the standard model (MSSM)~\cite{MSSM,MSSM1}
has drawn quite attention of physicists for a long time. Furthermore,
the matter-antimatter asymmetry in the universe requires that the baryon number
(B) must be broken. Meanwhile, the lepton number should also be broken and
as is well understood, existence of heavy Majorana neutrino(s) determines
tiny neutrino masses via the seesaw mechanism~\cite{seesaw,seesaw-1,seesaw-2,seesaw-3,seesaw-4}
and also naturally explains the  lepton number (L) violation.
Gauging baryon and leptons actually provides a natural framework for the seesaw
mechanism in the lepton sector, and the Peccei-Quinn mechanism solving
the strong CP problem in the quark sector~\cite{RFoot}. When the local B
and L gauge symmetries are broken around TeV scale, one does not need a
`desert region' between the weak and GUT scales to adequately suppress the
contribution of dimension 6 B-violating operators to proton decay \cite{BL-1}.
Note that gauging B-L symmetry does not address this issue since dimension 6
operators mentioned above are B-L invariant~\cite{PNath}.
Furthermore the simplest supersymmetric model with local $U(1)_{B-L}$ proposed
in Refs.~\cite{Perez1,Perez2,Perez3,Perez4} cannot account for LHC experimental data
of the 750 GeV resonance self-consistently unless we incorporate brand new
matter superfields~\cite{NSother-16}.
The authors of Refs.~\cite{BL_h,BL_h1} extended the MSSM by introducing
two extra U(1) gauge symmetries which correspond to baryon number B
and lepton number L as the BLMSSM, then in the new
theoretical framework they  investigated  decays of the SM-like
CP-even Higgs. Since the newly introduced quarks in BLMSSM are vector-like,
their masses can be well above $500\;{\rm GeV}$ without assuming a
large coupling to the Higgs doublets in this model. Therefore,
there does not exist a Landau pole for the Yukawa coupling~\cite{RFoot,BL-1,PNath}.
Additionally, the authors of Refs.~\cite{BL,BL-1,BL-1-1,BL-2,Feng,BL-3,BL-4,BL-5,BL-6,BL-7}
have done some studies in possible extension schemes of the SM where $U(1)_{_B}$ and
$U(1)_{_L}$ are spontaneously broken around ${\rm TeV}$ scale.

In this work, we explore the possibilities that the $750\;{\rm GeV}$ diphoton event
originates from decays of the CP-even scalars $h_{_B}^0,\;H_{_B}^0$ and/or the CP-odd scalar
$A_{_B}^0$ which are induced by spontaneously breaking the local $U(1)_{_B}$ symmetry.
These bosons are different from the CP-even Higgs $H^0$ and CP-odd Higgs $A^0$
which belongs to the $SU(2)$ doublet before spontaneously breaking.
Furthermore, since $h_{_B}^0,\;H_{_B}^0$ and $A_{_B}^0$ do not directly couple to the SM particles, so that
their decays into
SM particles can only be realized via loops which are suppressed by both the small couplings and the heavy intermediate agents (fermions or bosons).
In order to fit the well determined experimental data of the $125\;{\rm GeV}$ Higgs~\cite{ATLAS,CMS,CMS-ATLAS}, we
set the Yukawa coupling between the SM-like Higgs and exotic quarks into a suitable range,
and assume the Yukawa coupling between Higgs and exotic leptons to be negligible. Our numerical result indicates
that with a plausible parameter space,
the CP-odd scalar $A_{_B}^0$ in this model can naturally account for the experimental data on the $750\;{\rm GeV}$ excess measured by the
ATLAS and CMS collaborations.

Our work is organized as follows. In section \ref{sec2}, we briefly summarize the main ingredients of the BLMSSM,
then present the mass squared matrices for the neutral scalar sectors and the mass matrices for exotic quarks, respectively. We discuss
the decay widths for $\Phi\rightarrow\gamma\gamma,\;VV\;(V=Z,\;W,\;\Phi=h_{_B}^0,\;H_{_B}^0,\;A_{_B}^0)$
in section \ref{sec3}.
The numerical analyses are given in section \ref{sec4}, and eventually our summaries are made in the last section \ref{sec5}. Some formulae are collected in Appendixes \ref{app1}-\ref{LoopInduceCoupling}.

\section{A supersymmetric extension of the SM with B and L being local gauge symmetries\label{sec2}}
\indent\indent
When B and L are local gauge symmetries, one can enlarge the local gauge group of the SM to
$SU(3)_{_C}\otimes SU(2)_{_L}\otimes U(1)_{_Y}\otimes U(1)_{_B}\otimes U(1)_{_L}$.
In the supersymmetric extension of the SM proposed in Refs. \cite{BL_h,BL_h1}, the exotic superfields
include the new quarks $\hat{Q}_{_4}\sim(3,\;2,\;1/6,\;B_{_4},\;0)$,
$\hat{U}_{_4}^c\sim(\bar{3},\;1,\;-2/3,\;-B_{_4},\;0)$,
$\hat{D}_{_4}^c\sim(\bar{3},\;1,\;1/3,\;-B_{_4},\;0)$,
$\hat{Q}_{_5}^c\sim(\bar{3},\;2,\;-1/6,\;-(1+B_{_4}),\;0)$, $\hat{U}_{_5}\sim(3,\;1,\;2/3,\;1+B_{_4},\;0)$,
$\hat{D}_{_5}\sim(3,\;1,\;-1/3,\;1+B_{_4},\;0)$,
and the new leptons $\hat{L}_{_4}\sim(1,\;2,\;-1/2,\;0,\;L_{_4})$,
$\hat{E}_{_4}^c\sim(1,\;1,\;1,\;0,\;-L_{_4})$, $\hat{N}_{_4}^c\sim(1,\;1,\;0,\;0,\;-L_{_4})$,
$\hat{L}_{_5}^c\sim(1,\;2,\;1/2,\;0,\;-(3+L_{_4}))$, $\hat{E}_{_5}\sim(1,\;1,\;-1,\;0,\;3+L_{_4})$,
$\hat{N}_{_5}\sim(1,\;1,\;0,\;0,\;3+L_{_4})$ to cancel the $B$ and $L$ anomalies.
The `brand new' Higgs superfields $\hat{\Phi}_{_B}\sim(1,\;1,\;0,\;1,\;0)$ and
$\hat{\varphi}_{_B}\sim(1,\;1,\;0,\;-1,\;0)$ acquire nonzero vacuum expectation values (VEVs) to break
baryon number symmetry spontaneously. Meanwhile, nonzero VEVs of $\hat{\Phi}_{_B}$ and $\hat{\varphi}_{_B}$ also
induce large masses for the exotic quarks.
In addition, the superfields $\hat{S}_{_L}\sim(1,\;1,\;0,\;0,\;-2)$,
$\hat{\bar{S}}_{_L}\sim(1,\;1,\;0,\;0,\;2)$, $\hat{\Phi}_{_L}\sim(1,\;1,\;0,\;0,\;-3)$ and
$\hat{\varphi}_{_L}\sim(1,\;1,\;0,\;0,\;3)$ acquire nonzero VEVs to break lepton number symmetry spontaneously.
In addition, the VEVs of scalar components of $\hat{\Phi}_{_L}$ and $\hat{\varphi}_{_L}$
induce the TeV masses for 4th- and 5th-generation leptons, and the VEVs of scalar components of
$\hat{S}_{_L}$ and $\hat{\bar{S}}_{_L}$ produce the seesaw mechanism to result in tiny neutrino masses.
In order to avoid stability for the exotic quarks, the model also includes the
superfields $\hat{X}\sim(1,\;1,\;0,\;2/3+B_{_4},\;0)$ and
$\hat{X}^\prime\sim(1,\;1,\;0,\;-(2/3+B_{_4}),\;0)$. Actually, the lightest one can stand as a dark matter candidate.
The superpotential of the model is written as
\begin{eqnarray}
&&{\cal W}_{_{BLMSSM}}={\cal W}_{_{MSSM}}+{\cal W}_{_B}+{\cal W}_{_L}+{\cal W}_{_X}\;,
\label{superpotential1}
\end{eqnarray}
where ${\cal W}_{_{MSSM}}$ is the superpotential of the MSSM, and
\begin{eqnarray}
&&{\cal W}_{_B}=\lambda_{_Q}\hat{Q}_{_4}\hat{Q}_{_5}^c\hat{\Phi}_{_B}+\lambda_{_U}\hat{U}_{_4}^c\hat{U}_{_5}
\hat{\varphi}_{_B}+\lambda_{_D}\hat{D}_{_4}^c\hat{D}_{_5}\hat{\varphi}_{_B}+\mu_{_B}\hat{\Phi}_{_B}\hat{\varphi}_{_B}
\nonumber\\
&&\hspace{1.2cm}
+Y_{_{u_4}}\hat{Q}_{_4}\hat{H}_{_u}\hat{U}_{_4}^c+Y_{_{d_4}}\hat{Q}_{_4}\hat{H}_{_d}\hat{D}_{_4}^c
+Y_{_{u_5}}\hat{Q}_{_5}^c\hat{H}_{_d}\hat{U}_{_5}+Y_{_{d_5}}\hat{Q}_{_5}^c\hat{H}_{_u}\hat{D}_{_5}\;,
\nonumber\\
&&{\cal W}_{_L}=\lambda_{_L}\hat{L}_{_4}\hat{L}_{_5}^c\hat{\varphi}_{_L}+\lambda_{_E}\hat{E}_{_4}^c\hat{E}_{_5}
\hat{\Phi}_{_L}+\lambda_{_N}\hat{N}_{_4}^c\hat{N}_{_5}\hat{\Phi}_{_L}
+\mu_{_L}\hat{\Phi}_{_L}\hat{\varphi}_{_L}
\nonumber\\
&&\hspace{1.2cm}
+Y_{_{e_4}}\hat{L}_{_4}\hat{H}_{_d}\hat{E}_{_4}^c
+Y_{_{\nu_4}}\hat{L}_{_4}\hat{H}_{_u}\hat{N}_{_4}^c
+Y_{_{e_5}}\hat{L}_{_5}^c\hat{H}_{_u}\hat{E}_{_5}+Y_{_{\nu_5}}\hat{L}_{_5}^c\hat{H}_{_d}\hat{N}_{_5}
\nonumber\\
&&\hspace{1.2cm}
+Y_{_\nu}\hat{L}\hat{H}_{_u}\hat{N}^c
+\lambda_{_{N^c}}\hat{N}^c\hat{N}^c\hat{S}_{_L}+\mu_{_S}\hat{S}_{_L}\hat{\bar{S}}_{_L}\;,
\nonumber\\
&&{\cal W}_{_X}=\lambda_1\hat{Q}\hat{Q}_{_5}^c\hat{X}+\lambda_2\hat{U}^c\hat{U}_{_5}\hat{X}^\prime
+\lambda_3\hat{D}^c\hat{D}_{_5}\hat{X}^\prime+\mu_{_X}\hat{X}\hat{X}^\prime\;.
\label{superpotential-BL}
\end{eqnarray}
In the superpotential given above, the exotic quarks obtain ${\rm TeV}$ scale masses after $\Phi_{_B},\;\varphi_{_B}$
acquire nonzero VEVs.
Correspondingly, the soft breaking terms are generally given as
\begin{eqnarray}
&&{\cal L}_{_{soft}}={\cal L}_{_{soft}}^{MSSM}-(m_{_{\tilde{N}^c}}^2)_{_{IJ}}\tilde{N}_I^{c*}\tilde{N}_J^c
-m_{_{\tilde{Q}_4}}^2\tilde{Q}_{_4}^\dagger\tilde{Q}_{_4}-m_{_{\tilde{U}_4}}^2\tilde{U}_{_4}^{c*}\tilde{U}_{_4}^c
-m_{_{\tilde{D}_4}}^2\tilde{D}_{_4}^{c*}\tilde{D}_{_4}^c
\nonumber\\
&&\hspace{1.3cm}
-m_{_{\tilde{Q}_5}}^2\tilde{Q}_{_5}^{c\dagger}\tilde{Q}_{_5}^c-m_{_{\tilde{U}_5}}^2\tilde{U}_{_5}^*\tilde{U}_{_5}
-m_{_{\tilde{D}_5}}^2\tilde{D}_{_5}^*\tilde{D}_{_5}-m_{_{\tilde{L}_4}}^2\tilde{L}_{_4}^\dagger\tilde{L}_{_4}
-m_{_{\tilde{\nu}_4}}^2\tilde{\nu}_{_4}^{c*}\tilde{\nu}_{_4}^c
\nonumber\\
&&\hspace{1.3cm}
-m_{_{\tilde{E}_4}}^2\tilde{e}_{_4}^{c*}\tilde{e}_{_4}^c-m_{_{\tilde{L}_5}}^2\tilde{L}_{_5}^{c\dagger}\tilde{L}_{_5}^c
-m_{_{\tilde{\nu}_5}}^2\tilde{\nu}_{_5}^*\tilde{\nu}_{_5}-m_{_{\tilde{E}_5}}^2\tilde{e}_{_5}^*\tilde{e}_{_5}
-m_{_{\Phi_{_B}}}^2\Phi_{_B}^*\Phi_{_B}
\nonumber\\
&&\hspace{1.3cm}
-m_{_{\varphi_{_B}}}^2\varphi_{_B}^*\varphi_{_B}-m_{_{\Phi_{_L}}}^2\Phi_{_L}^*\Phi_{_L}
-m_{_{\varphi_{_L}}}^2\varphi_{_L}^*\varphi_{_L}-\Big(m_{_B}\lambda_{_B}\lambda_{_B}
+m_{_L}\lambda_{_L}\lambda_{_L}+h.c.\Big)
\nonumber\\
&&\hspace{1.3cm}
+\Big\{A_{_{u_4}}Y_{_{u_4}}\tilde{Q}_{_4}H_{_u}\tilde{U}_{_4}^c+A_{_{d_4}}Y_{_{d_4}}\tilde{Q}_{_4}H_{_d}\tilde{D}_{_4}^c
+A_{_{u_5}}Y_{_{u_5}}\tilde{Q}_{_5}^cH_{_d}\tilde{U}_{_5}+A_{_{d_5}}Y_{_{d_5}}\tilde{Q}_{_5}^cH_{_u}\tilde{D}_{_5}
\nonumber\\
&&\hspace{1.3cm}
+A_{_{BQ}}\lambda_{_Q}\tilde{Q}_{_4}\tilde{Q}_{_5}^c\Phi_{_B}+A_{_{BU}}\lambda_{_U}\tilde{U}_{_4}^c\tilde{U}_{_5}\varphi_{_B}
+A_{_{BD}}\lambda_{_D}\tilde{D}_{_4}^c\tilde{D}_{_5}\varphi_{_B}+B_{_B}\mu_{_B}\Phi_{_B}\varphi_{_B}
+h.c.\Big\}
\nonumber\\
&&\hspace{1.3cm}
+\Big\{A_{_{e_4}}Y_{_{e_4}}\tilde{L}_{_4}H_{_d}\tilde{E}_{_4}^c+A_{_{N_4}}Y_{_{N_4}}\tilde{L}_{_4}H_{_u}\tilde{N}_{_4}^c
+A_{_{e_5}}Y_{_{e_5}}\tilde{L}_{_5}^cH_{_u}\tilde{E}_{_5}+A_{_{N_5}}Y_{_{\nu_5}}\tilde{L}_{_5}^cH_{_d}\tilde{N}_{_5}
\nonumber\\
&&\hspace{1.3cm}
+A_{_N}Y_{_N}\tilde{L}H_{_u}\tilde{N}^c+A_{_{LL}}\lambda_{_L}\tilde{L}_{_4}\tilde{L}_{_5}^c\varphi_{_L}
+A_{_{LE}}\lambda_{_E}\tilde{E}_{_4}^c\tilde{E}_{_5}\Phi_{_L}
+A_{_{LN}}\lambda_{_N}\tilde{N}_{_4}^c\tilde{N}_{_5}\Phi_{_L}
\nonumber\\
&&\hspace{1.3cm}
+B_{_L}\mu_{_L}\Phi_{_L}\varphi_{_L}
+A_{_{N^c}}\lambda_{_{N^c}}\tilde{N}^c\tilde{N}^cS_{_L}
+B_{_S}\mu_{_S}S_{_L}\bar{S}_{_L}+h.c.\Big\}
\nonumber\\
&&\hspace{1.3cm}
+\Big\{A_1\lambda_1\tilde{Q}\tilde{Q}_{_5}^cX+A_2\lambda_2\tilde{U}^c\tilde{U}_{_5}X^\prime
+A_3\lambda_3\tilde{D}^c\tilde{D}_{_5}X^\prime+B_{_X}\mu_{_X}XX^\prime+h.c.\Big\}\;,
\label{soft-breaking}
\end{eqnarray}
where ${\cal L}_{_{soft}}^{MSSM}$ is the soft breaking terms for the MSSM, $\lambda_B,\;\lambda_L$
are gauginos of $U(1)_{_B}$ and $U(1)_{_L}$, respectively.
After the $SU(2)_L$ doublets $H_{_u},\;H_{_d}$ and $SU(2)_L$ singlets $\Phi_{_B},\;\varphi_{_B},\;\Phi_{_L},\;
\varphi_{_L}$ acquire nonzero VEVs: $\upsilon_{_u},\;\upsilon_{_d},\;\upsilon_{_{B}},\;\overline{\upsilon}_{_{B}}$,
and $\upsilon_{_L},\;\overline{\upsilon}_{_L}$, then we have
\begin{eqnarray}
&&H_{_u}=\left(\begin{array}{c}H_{_u}^+\\{1\over\sqrt{2}}\Big(\upsilon_{_u}+H_{_u}^0+iP_{_u}^0\Big)\end{array}\right)\;,
\nonumber\\
&&H_{_d}=\left(\begin{array}{c}{1\over\sqrt{2}}\Big(\upsilon_{_d}+H_{_d}^0+iP_{_d}^0\Big)\\H_{_d}^-\end{array}\right)\;,
\nonumber\\
&&\Phi_{_B}={1\over\sqrt{2}}\Big(\upsilon_{_B}+\Phi_{_B}^0+iP_{_B}^0\Big)\;,
\nonumber\\
&&\varphi_{_B}={1\over\sqrt{2}}\Big(\overline{\upsilon}_{_B}+\varphi_{_B}^0+i\overline{P}_{_B}^0\Big)\;,
\nonumber\\
&&\Phi_{_L}={1\over\sqrt{2}}\Big(\upsilon_{_L}+\Phi_{_L}^0+iP_{_L}^0\Big)\;,
\nonumber\\
&&\varphi_{_L}={1\over\sqrt{2}}\Big(\overline{\upsilon}_{_L}+\varphi_{_L}^0+i\overline{P}_{_L}^0\Big)\;,
\label{VEVs}
\end{eqnarray}
and the local gauge symmetry $SU(2)_{_L}\otimes U(1)_{_Y}\otimes U(1)_{_B}\otimes U(1)_{_L}$ is broken
down to the electromagnetic symmetry $U(1)_{{em}}$, where
\begin{eqnarray}
&&G^\pm=\cos\beta H_{_d}^\pm+\sin\beta H_{_u}^\pm\;,
\label{Goldstone1}
\end{eqnarray}
denotes the charged Goldstone boson, and
\begin{eqnarray}
&&G^0=\cos\beta P_{_d}^0+\sin\beta P_{_u}^0\;,\nonumber\\
&&G_{_B}^0=\cos\beta_{_B} P_{_B}^0+\sin\beta_{_B}\overline{P}_{_B}^0\;,\nonumber\\
&&G_{_L}^0=\cos\beta_{_L} P_{_L}^0+\sin\beta_{_L}\overline{P}_{_L}^0\;,
\label{Goldstone2}
\end{eqnarray}
denote the neutral Goldstone bosons, respectively. Here $\tan\beta=\upsilon_{_u}/\upsilon_{_d},\;
\tan\beta_{_B}=\overline{\upsilon}_{_B}/\upsilon_{_B}$ and $\tan\beta_{_L}=\overline{\upsilon}_{_L}/\upsilon_{_L}$.
Correspondingly, the physical neutral pseudoscalar fields are
\begin{eqnarray}
&&A^0=-\sin\beta P_{_d}^0+\cos\beta P_{_u}^0\;,\nonumber\\
&&A_{_B}^0=-\sin\beta_{_B} P_{_B}^0+\cos\beta_{_B}\overline{P}_{_B}^0\;,\nonumber\\
&&A_{_L}^0=-\sin\beta_{_L} P_{_L}^0+\cos\beta_{_L}\overline{P}_{_L}^0\;.
\label{neutral-pseudoscalar}
\end{eqnarray}
At the tree level, the masses for those particles are respectively formulated as
\begin{eqnarray}
&&m_{_{A^0}}^2={B\mu\over\cos\beta\sin\beta}\;,\nonumber\\
&&m_{_{A_{_B}^0}}^2={B_{_B}\mu_{_B}\over\cos\beta_{_B}\sin\beta_{_B}}\;,\nonumber\\
&&m_{_{A_{_L}^0}}^2={B_{_L}\mu_{_L}\over\cos\beta_{_L}\sin\beta_{_L}}\;.
\label{pseudoscalar-mass}
\end{eqnarray}
Meanwhile the charged Higgs is
\begin{eqnarray}
&&H^\pm=-\sin\beta H_{_d}^\pm+\cos\beta H_{_u}^\pm\; ,
\label{charged-Higgs}
\end{eqnarray}
with the tree level mass square
\begin{eqnarray}
&&m_{_{H^\pm}}^2=m_{_{A^0}}^2+m_{_{\rm W}}^2\;.
\label{charged-mass}
\end{eqnarray}
In the two Higgs doublet sector, the mass square matrix of neutral CP-even Higgs is diagonalized by a rotation
\begin{eqnarray}
&&\left(\begin{array}{l}H^0\\h^0\end{array}\right)=\left(\begin{array}{cc}\cos\alpha&\sin\alpha\\
-\sin\alpha&\cos\alpha\end{array}\right)\left(\begin{array}{l}H_{_d}^0\\H_{_u}^0\end{array}\right)\;,
\label{CP-even-Higgs}
\end{eqnarray}
where $h^0$ is the lightest neutral CP-even Higgs.

In the basis $(\Phi_{_B}^0,\;\varphi_{_B}^0)$, the mass square matrix is
\begin{eqnarray}
&&{\cal M}_{_{EB}}^2=\left(\begin{array}{ll}m_{_{Z_B}}^2\cos^2\beta_{_B}+m_{_{A_{_B}^0}}^2\sin^2\beta_{_B},\;&
(m_{_{Z_B}}^2+m_{_{A_{_B}^0}}^2)\cos\beta_{_B}\sin\beta_{_B}\\
(m_{_{Z_B}}^2+m_{_{A_{_B}^0}}^2)\cos\beta_{_B}\sin\beta_{_B},\;&
m_{_{Z_B}}^2\sin^2\beta_{_B}+m_{_{A_{_B}^0}}^2\cos^2\beta_{_B}
\end{array}\right)\;,
\label{CPevenB-mass}
\end{eqnarray}
where $m_{_{Z_B}}^2=g_{_B}^2(\upsilon_{_B}^2+\overline{\upsilon}_{_B}^2)$ is mass square of
the neutral $U(1)_{_B}$ gauge boson $Z_{_B}$. Defining the mixing angle $\alpha_{_B}$ through
\begin{eqnarray}
&&\tan2\alpha_{_B}={m_{_{Z_B}}^2+m_{_{A_{_B}^0}}^2\over m_{_{Z_B}}^2-m_{_{A_{_B}^0}}^2}
\tan2\beta_{_B}\;,
\label{Mixing-B}
\end{eqnarray}
we obtain two mass eigenstates as
\begin{eqnarray}
&&\left(\begin{array}{l}H_{_B}^0\\ h_{_B}^0\end{array}\right)=
\left(\begin{array}{cc}\cos\alpha_{_B}&\sin\alpha_{_B}\\-\sin\alpha_{_B}&\cos\alpha_{_B}\end{array}\right)
\left(\begin{array}{l}\Phi_{_B}^0\\ \varphi_{_B}^0\end{array}\right)\;.
\label{Mass-eigenstates-B}
\end{eqnarray}

Similarly the mass square matrix for $(\Phi_{_L}^0,\;\varphi_{_L}^0)$
is written as
\begin{eqnarray}
&&{\cal M}_{_{EL}}^2=\left(\begin{array}{ll}m_{_{Z_L}}^2\cos^2\beta_{_L}+m_{_{A_{_L}^0}}^2\sin^2\beta_{_L},\;&
(m_{_{Z_L}}^2+m_{_{A_{_L}^0}}^2)\cos\beta_{_L}\sin\beta_{_L}\\
(m_{_{Z_L}}^2+m_{_{A_{_L}^0}}^2)\cos\beta_{_L}\sin\beta_{_L},\;&
m_{_{Z_L}}^2\sin^2\beta_{_L}+m_{_{A_{_L}^0}}^2\cos^2\beta_{_L}
\end{array}\right)\;,
\label{CPevenL-mass}
\end{eqnarray}
with $m_{_{Z_L}}^2=4g_{_L}^2(\upsilon_{_L}^2+\overline{\upsilon}_{_L}^2)$ denoting mass square of
the neutral $U(1)_{_L}$ gauge boson $Z_{_L}$.
We can obtain two mass eigenstates as
\begin{eqnarray}
&&\left(\begin{array}{l}H_{_L}^0\\ h_{_L}^0\end{array}\right)=
\left(\begin{array}{cc}\cos\alpha_{_L}&\sin\alpha_{_L}\\-\sin\alpha_{_L}&\cos\alpha_{_L}\end{array}\right)
\left(\begin{array}{l}\Phi_{_L}^0\\ \varphi_{_L}^0\end{array}\right)\;.
\label{Mass-eigenstates-L}
\end{eqnarray}

The mass matrix for the exotic quarks of charge $2/3$ which are four-component Dirac spinors  is
\begin{eqnarray}
&&-{\cal L}_{_{t^\prime}}^{mass}=\left(\begin{array}{ll}\bar{t}_{_{4R}}^\prime,&\bar{t}_{_{5R}}^\prime\end{array}\right)
\left(\begin{array}{ll}{1\over\sqrt{2}}\lambda_{_Q}\upsilon_{_B},&-{1\over\sqrt{2}}Y_{_{u_5}}\upsilon_{_d}\\
-{1\over\sqrt{2}}Y_{_{u_4}}\upsilon_{_u},&{1\over\sqrt{2}}\lambda_{_U}\overline{\upsilon}_{_B}
\end{array}\right)\left(\begin{array}{l}t_{_{4L}}^\prime\\t_{_{5L}}^\prime\end{array}\right)+h.c.
\label{Qmass-matrix-2/3}
\end{eqnarray}
Performing unitary transformations $U_{_{t}}$ and $W_{_{t}} $
\begin{eqnarray}
&&\left(\begin{array}{l}t_{_{4L}}\\t_{_{5L}}\end{array}\right)
=U_{_{t}}^\dagger\cdot\left(\begin{array}{l}t_{_{4L}}^\prime\\t_{_{5L}}^\prime\end{array}\right)\;,\;\;
\left(\begin{array}{l}t_{_{4R}}\\t_{_{5R}}\end{array}\right)
=W_{_{t}}^\dagger\cdot\left(\begin{array}{l}t_{_{4R}}^\prime\\t_{_{5R}}^\prime\end{array}\right)\;,
\label{Qmixing-2/3-a}
\end{eqnarray}
we diagonalize the mass matrix for the vector quarks of charge $2/3$:
\begin{eqnarray}
&&W_{_{t}}^\dagger\cdot\left(\begin{array}{ll}{1\over\sqrt{2}}\lambda_{_Q}\upsilon_{_B},&
-{1\over\sqrt{2}}Y_{_{u_5}}\upsilon_{_d}\\-{1\over\sqrt{2}}Y_{_{u_4}}\upsilon_{_u},&
{1\over\sqrt{2}}\lambda_{_U}\overline{\upsilon}_{_B}\end{array}\right)\cdot U_{_{t}}
={\it diag}\Big(m_{_{t_4}},\;m_{_{t_5}}\Big).
\label{Qmixing-2/3-b}
\end{eqnarray}
Similarly we write the mass matrix for the exotic quarks of charge $-1/3$ as
\begin{eqnarray}
&&-{\cal L}_{_{b^\prime}}^{mass}=\left(\begin{array}{ll}\bar{b}_{_{4R}}^\prime,&\bar{b}_{_{5R}}^\prime\end{array}\right)
\left(\begin{array}{ll}-{1\over\sqrt{2}}\lambda_{_Q}\upsilon_{_B},&-{1\over\sqrt{2}}Y_{_{d_5}}\upsilon_{_u}\\
-{1\over\sqrt{2}}Y_{_{d_4}}\upsilon_{_d},&{1\over\sqrt{2}}\lambda_{_D}\overline{\upsilon}_{_B}
\end{array}\right)\left(\begin{array}{l}b_{_{4L}}^\prime\\b_{_{5L}}^\prime\end{array}\right)+h.c.
\label{Qmass-matrix-1/3}
\end{eqnarray}
Adopting unitary transformations
\begin{eqnarray}
&&\left(\begin{array}{l}b_{_{4L}}\\b_{_{5L}}\end{array}\right)
=U_{_{b}}^\dagger\cdot\left(\begin{array}{l}b_{_{4L}}^\prime\\b_{_{5L}}^\prime\end{array}\right)\;,\;\;
\left(\begin{array}{l}b_{_{4R}}\\b_{_{5R}}\end{array}\right)
=W_{_{b}}^\dagger\cdot\left(\begin{array}{l}b_{_{4R}}^\prime\\b_{_{5R}}^\prime\end{array}\right)\;,
\label{Qmixing-1/3-a}
\end{eqnarray}
one can diagonalize the  mass matrix for the vector quarks of charge $-1/3$ as
\begin{eqnarray}
&&W_{_{b}}^\dagger\cdot\left(\begin{array}{ll}-{1\over\sqrt{2}}\lambda_{_Q}\upsilon_{_B},&
-{1\over\sqrt{2}}Y_{_{d_5}}\upsilon_{_u}\\-{1\over\sqrt{2}}Y_{_{d_4}}\upsilon_{_d},&
{1\over\sqrt{2}}\lambda_{_D}\overline{\upsilon}_{_B}\end{array}\right)\cdot U_{_{b}}
={\it diag}\Big(m_{_{b_4}},\;m_{_{b_5}}\Big)\;.
\label{Qmixing-1/3-b}
\end{eqnarray}

Using the superpotential in Eq.~(\ref{superpotential1}) and introducing the soft breaking terms,
we write the mass square matrices for the exotic scalar quarks as
\begin{eqnarray}
&&-{\cal L}_{_{\widetilde{EQ}}}^{mass}=\tilde{t}^{\prime\dagger}\cdot
{\cal M}_{\tilde{t}^\prime}^2\cdot\tilde{t}^\prime
+\tilde{b}^{\prime\dagger}\cdot {\cal M}_{\tilde{b}^\prime}^2\cdot\tilde{b}^\prime
\label{SQmass-2/3}
\end{eqnarray}
with $\tilde{t}^{\prime T}=(\tilde{Q}_{_4}^{1},\;\tilde{U}_{_4}^{c*},\;\tilde{Q}_{_5}^{2c*},\;
\tilde{U}_{_5})$, $\tilde{b}^{\prime T}=(\tilde{Q}_{_4}^{2},\;\tilde{D}_{_4}^{c*},
\;\tilde{Q}_{_5}^{1c*},\;\tilde{D}_{_5}^{*})$.
The concrete expressions for the $4\times4$
mass square matrices ${\cal M}_{\tilde{t}^\prime}^2,\;{\cal M}_{\tilde{b}^\prime}^2$,
and the couplings between the neutral  Higgs and exotic scalar quarks are collected
elsewhere~\cite{Feng}, the couplings
between heavy neutral Higgs and exotic quarks can also be found there.

The mass matrix for exotic neutrinos which are four-component spinors, is
\begin{eqnarray}
&&-{\cal L}_{_{\nu^\prime}}^{mass}=\left(\begin{array}{ll}\bar{\nu}_{_{4R}}^\prime,&\bar{\nu}_{_{5R}}^\prime\end{array}\right)
\left(\begin{array}{ll}{1\over\sqrt{2}}\lambda_{_L}\overline{\upsilon}_{_L},&-{1\over\sqrt{2}}Y_{_{\nu_5}}\upsilon_{_d}\\
-{1\over\sqrt{2}}Y_{_{\nu_4}}\upsilon_{_u},&{1\over\sqrt{2}}\lambda_{_N}\upsilon_{_L}
\end{array}\right)\left(\begin{array}{l}\nu_{_{4L}}^\prime\\\nu_{_{5L}}^\prime\end{array}\right)+h.c.
\label{Nmass-matrix}
\end{eqnarray}
Similarly the mass matrix for exotic charged leptons is
\begin{eqnarray}
&&-{\cal L}_{_{e^\prime}}^{mass}=\left(\begin{array}{ll}\bar{e}_{_{4R}}^\prime,&\bar{e}_{_{5R}}^\prime\end{array}\right)
\left(\begin{array}{ll}-{1\over\sqrt{2}}\lambda_{_L}\overline{\upsilon}_{_L},&-{1\over\sqrt{2}}Y_{_{e_5}}\upsilon_{_u}\\
-{1\over\sqrt{2}}Y_{_{e_4}}\upsilon_{_d},&{1\over\sqrt{2}}\lambda_{_E}\upsilon_{_L}
\end{array}\right)\left(\begin{array}{l}e_{_{4L}}^\prime\\e_{_{5L}}^\prime\end{array}\right)+h.c.
\label{Emass-matrix}
\end{eqnarray}

Including those `new' particles mentioned above, the evaluations of gauge couplings
are described by the renormalization group equations (RGEs)~\cite{T.P.Cheng,P.Langacker}
\begin{eqnarray}
&&{dg_{_i}\over dt}={1\over2}\beta_{g_{_i}},\;(i=1,\;2,\;3),
\nonumber\\
&&{dg_{_B}\over dt}={1\over2}\beta_{g_{_B}},\;{dg_{_L}\over dt}={1\over2}\beta_{g_{_L}}\;,
\label{RGE1}
\end{eqnarray}
where $t=\ln Q^2$.
Adopting the step approximation in contributions from new particles to the
$\beta$ functions~\cite{R.Hempfling}, we then find
\begin{eqnarray}
&&\beta_{g_{_3}}=-{g_{_3}^3\over16\pi^2}\Big\{\Big(11-{10\over3}-{2\over3}\theta_{_t}\Big)
-{1\over3}\sum\limits_{\alpha=4}^5\Big(2\theta_{_{Q_\alpha}}+\theta_{_{U_\alpha}}+\theta_{_{D_\alpha}}\Big)
-2\theta_{_{\tilde g}}
\nonumber\\
&&\hspace{1.2cm}
-{1\over24}\sum\limits_{i=1}^5\Big(2\theta_{_{\tilde{Q}_i}}
+\theta_{_{\tilde{U}_i}}+\theta_{_{\tilde{D}_i}}\Big)\Big\}\;,
\nonumber\\
&&\beta_{g_{_2}}=-{g_{_2}^3\over16\pi^2}\Big\{\Big({22\over3}-3-\theta_{_t}\Big)
-{1\over3}\theta_{_{A^0}}-{4\over3}\theta_{_{\tilde W}}-{2\over3}\theta_{_{\tilde H}}
-{1\over3}\sum\limits_{\alpha=4}^5\Big(\theta_{_{L_\alpha}}+3\theta_{_{Q_\alpha}}\Big)
\nonumber\\
&&\hspace{1.2cm}
-{1\over24}\sum\limits_{i=1}^5\Big(\theta_{_{\tilde{L}_i}}
+3\theta_{_{\tilde{Q}_i}}\Big)\Big\}\;,
\nonumber\\
&&\beta_{g_{_1}}={g_{_1}^3\over16\pi^2}\Big\{\Big({51\over9}+\theta_{_t}\Big)
+{1\over3}\theta_{_{A^0}}+{2\over3}\theta_{_{\tilde H}}
+{4\over3}\sum\limits_{\alpha=4}^5\Big({1\over4}\theta_{_{L_\alpha}}+{1\over2}\theta_{_{E_\alpha}}
+{1\over12}\theta_{_{Q_\alpha}}+{2\over3}\theta_{_{U_\alpha}}
\nonumber\\
&&\hspace{1.2cm}
+{1\over6}\theta_{_{D_\alpha}}\Big)
+{1\over6}\sum\limits_{i=1}^5\Big({1\over4}\theta_{_{\tilde{L}_i}}+{1\over2}\theta_{_{\tilde{E}_i}}
+{1\over12}\theta_{_{\tilde{Q}_i}}+{2\over3}\theta_{_{\tilde{U}_i}}
+{1\over6}\theta_{_{\tilde{D}_i}}\Big)\Big\}\;,
\nonumber\\
&&\beta_{g_{_B}}={g_{_B}^3\over16\pi^2}\Big\{\Big(2+{2\over3}\theta_{_t}\Big)
+{1\over36}\sum\limits_{i=1}^3\Big(2\theta_{_{\tilde{Q}_i}}+\theta_{_{\tilde{U}_i}}
+\theta_{_{\tilde{D}_i}}\Big)+{1\over3}\Big(2\theta_{_{\tilde{\Phi}_B}}
+2\theta_{_{\tilde{\phi}_B}}
\nonumber\\
&&\hspace{1.2cm}
+{1\over4}\theta_{_{\Phi_B}}+{1\over4}\theta_{_{\phi_B}}\Big)+({1\over3}+{B_{_4}\over2})^2
\Big(2\theta_{_{\tilde{X}}}+2\theta_{_{\tilde{X}^\prime}}+{1\over4}\theta_{_{X}}+{1\over4}\theta_{_{X^\prime}}\Big)
\nonumber\\
&&\hspace{1.2cm}
+B_{_4}^2\Big(4\theta_{_{Q_4}}+2\theta_{_{U_4}}+2\theta_{_{D_4}}+{1\over2}\theta_{_{\tilde{Q}_4}}
+{1\over4}\theta_{_{\tilde{U}_4}}+{1\over4}\theta_{_{\tilde{D}_4}}\Big)
\nonumber\\
&&\hspace{1.2cm}
+(1+B_{_4})^2\Big(4\theta_{_{Q_5}}+2\theta_{_{U_5}}+2\theta_{_{D_5}}+{1\over2}\theta_{_{\tilde{Q}_5}}
+{1\over4}\theta_{_{\tilde{U}_5}}+{1\over4}\theta_{_{\tilde{D}_5}}\Big)\Big\}\;,
\nonumber\\
&&\beta_{g_{_L}}={g_{_L}^3\over16\pi^2}\Big\{6+{2\over3}\sum\limits_{i=1}^3\theta_{_{N_i}}
+{1\over12}\sum\limits_{i=1}^3\Big(2\theta_{_{\tilde{L}_i}}+\theta_{_{\tilde{E}_i}}
+\theta_{_{\tilde{N}_i}}\Big)+{4\over3}\Big(2\theta_{_{\tilde{S}_L}}
+2\theta_{_{\tilde{\bar{S}}_L}}
\nonumber\\
&&\hspace{1.2cm}
+{1\over4}\theta_{_{S_L}}+{1\over4}\theta_{_{\bar{S}_L}}\Big)+3\Big(2\theta_{_{\tilde{\Phi}_L}}
+2\theta_{_{\tilde{\phi}_L}}+{1\over4}\theta_{_{\Phi_L}}+{1\over4}\theta_{_{\phi_L}}\Big)
\nonumber\\
&&\hspace{1.2cm}
+{L_{_4}^2\over3}\Big(4\theta_{_{L_4}}+2\theta_{_{N_4}}+2\theta_{_{E_4}}+{1\over2}\theta_{_{\tilde{L}_4}}
+{1\over4}\theta_{_{\tilde{N}_4}}+{1\over4}\theta_{_{\tilde{E}_4}}\Big)
\nonumber\\
&&\hspace{1.2cm}
+{(1+L_{_4})^2\over3}\Big(4\theta_{_{L_5}}+2\theta_{_{N_5}}+2\theta_{_{E_5}}+{1\over2}\theta_{_{\tilde{L}_5}}
+{1\over4}\theta_{_{\tilde{N}_5}}+{1\over4}\theta_{_{\tilde{E}_5}}\Big)\Big\}\;,
\label{beta-functions}
\end{eqnarray}
with
\begin{eqnarray}
&&\theta_{_a}=\theta(\ln{Q^2\over m_{_a}^2})=\left\{
\begin{array}{l}
1\;\;\;{\rm for}\: Q> m_{_a}\;,\\
0\;\;\;{\rm for}\: Q\leq m_{_a}\;.
\end{array}\right.
\label{theta-function}
\end{eqnarray}
To simplify our discussion below, we assume
new particles with masses of roughly same order $\Lambda_{_{NP}}$. Using the evolution
equations in Eq.~(\ref{RGE1}), we obtain the effective couplings for
$\alpha_{_i}\;(i=3,\;,2,\;1)$ as
\begin{eqnarray}
&&\alpha_{_3}(\Lambda)=\left\{\begin{array}{l}
{\alpha_{_3}(m_{_{\rm Z}})\over1+{23\over3}{\alpha_{_3}(m_{_{\rm Z}})\over4\pi}
\ln{\Lambda^2\over m_{_{\rm Z}}^2}}\;,\;m_{_{\rm Z}}<\Lambda\le m_{_t}\\
{\alpha_{_3}(m_{_t})\over1+7{\alpha_{_3}(m_{_t})\over4\pi}
\ln{\Lambda^2\over m_{_t}^2}}\;,\;m_{_t}<\Lambda\le \Lambda_{_{NP}}\\
{\alpha_{_3}(\Lambda_{_{NP}})\over1+{3\over2}{\alpha_{_3}(\Lambda_{_{NP}})\over4\pi}
\ln{\Lambda^2\over \Lambda_{_{NP}}^2}}\;,\;\Lambda>\Lambda_{_{NP}}\\
\end{array}\right.\;,
\nonumber\\
&&\alpha_{_2}(\Lambda)=\left\{\begin{array}{l}
{\alpha_{_2}(m_{_{\rm Z}})\over1+{13\over3}{\alpha_{_2}(m_{_{\rm Z}})\over4\pi}
\ln{\Lambda^2\over m_{_{\rm Z}}^2}}\;,\;m_{_{\rm Z}}<\Lambda\le m_{_t}\\
{\alpha_{_2}(m_{_t})\over1+\frac{10}{3}{\alpha_{_2}(m_{_t})\over4\pi}
\ln{\Lambda^2\over m_{_t}^2}}\;,\;m_{_t}<\Lambda\le \Lambda_{_{NP}}\\
{\alpha_{_2}(\Lambda_{_{NP}})\over1-{5\over2}{\alpha_{_2}(\Lambda_{_{NP}})\over4\pi}
\ln{\Lambda^2\over \Lambda_{_{NP}}^2}}\;,\;\Lambda>\Lambda_{_{NP}}\\
\end{array}\right.\;,
\nonumber\\
&&\alpha_{_1}(\Lambda)=\left\{\begin{array}{l}
{\alpha_{_1}(m_{_{\rm Z}})\over1-{51\over9}{\alpha_{_1}(m_{_{\rm Z}})\over4\pi}
\ln{\Lambda^2\over m_{_{\rm Z}}^2}}\;,\;m_{_{\rm Z}}<\Lambda\le m_{_t}\\
{\alpha_{_1}(m_{_t})\over1-\frac{20}{3}{\alpha_{_1}(m_{_t})\over4\pi}
\ln{\Lambda^2\over m_{_t}^2}}\;,\;m_{_t}<\Lambda\le \Lambda_{_{NP}}\\
{\alpha_{_1}(\Lambda_{_{NP}})\over1-{27\over2}{\alpha_{_1}(\Lambda_{_{NP}})\over4\pi}
\ln{\Lambda^2\over \Lambda_{_{NP}}^2}}\;,\;\Lambda>\Lambda_{_{NP}}\\
\end{array}\right.\;,
\label{effective-couplings1}
\end{eqnarray}
with $\alpha_{_i}=g_{_i}^2/(4\pi)$. Obviously there is no Landau singularity
in the strong interaction coupling $\alpha_{_3}$ as $\Lambda>\Lambda_{_{NP}}$, the Landau singularities
for $\alpha_{_{1,2}}$ are approached as
\begin{eqnarray}
&&\Lambda_{_{LS}}^{(1)}\simeq\Lambda_{_{NP}}\exp\Big[{4\pi\over27\alpha_{_1}(\Lambda_{_{NP}})}\Big]
\;,\nonumber\\
&&\Lambda_{_{LS}}^{(2)}\simeq\Lambda_{_{NP}}\exp\Big[{4\pi\over5\alpha_{_2}(\Lambda_{_{NP}})}\Big]\;.
\label{Landau-singularities}
\end{eqnarray}
Choosing $\alpha(m_{_{\rm Z}})=1/128,\;s_{_{\rm W}}^2(m_{_{\rm Z}})=0.23$,
$m_{_t}=174$ GeV, $m_{_{\rm Z}}=91.19$ GeV, $\Lambda_{_{NP}}=3$ TeV, we get
$\Lambda_{_{LS}}^{(1)}\simeq4.7\times 10^{22}$ GeV and $\Lambda_{_{LS}}^{(2)}\simeq5.6 \times10^{37}$ GeV which is above the Planck
scale $\Lambda_{_{Planck}}\sim10^{19}$ GeV, so $\alpha_{_1}$ and $\alpha_{_2}$ is safe, i.e. not bothered by
the singularity.

Furthermore, the Landau singularities
of $\alpha_{_{B,L}}$ are written as
\begin{eqnarray}
&&\Lambda_{_{LS}}^{(B)}\simeq\Lambda_{_{NP}}\exp\Big[{2\pi\over b_{_B}\alpha_{_B}(\Lambda_{_{NP}})}\Big]
\;,\nonumber\\
&&\Lambda_{_{LS}}^{(L)}\simeq\Lambda_{_{NP}}\exp\Big[{2\pi\over b_{_L}\alpha_{_L}(\Lambda_{_{NP}})}\Big]\;,
\label{Landau-singularitiesBL}
\end{eqnarray}
with
\begin{eqnarray}
&&b_{_B}={9\over2}\Big[1+({1\over3}+B_{_4})^2+2B_{_4}^2+2(1+B_{_4})^2\Big],
\nonumber\\
&&b_{_L}={57\over2}+3L_{_4}^2+3(1+L_{_4})^2.
\end{eqnarray}
Choose $B_{_4}=L_{_4}=0$, $\Lambda_{_{NP}}=3$ TeV, $g_{_B}(\Lambda_{_{NP}})
=0.35$ and $g_{_L}(\Lambda_{_{NP}})=0.2$, one obtains $\Lambda_{_{LS}}^{(B)}\simeq3.0 \times10^{23}$ GeV,
$\Lambda_{_{LS}}^{(L)}\simeq 4.9 \times10^{30}$ GeV, respectively.

\section{$gg\rightarrow \Phi$ and $\Phi\rightarrow\gamma\gamma,\;ZZ,\;Z\gamma,\;
WW,\;t\bar{t}$\label{sec3}}
\indent\indent
It is well known for quite some while that radiative corrections modify the tree level mass square
matrix of neutral Higgs substantially in the MSSM, where the main effect originates from one-loop diagrams involving the top
quark and its scalar partners $\tilde{t}_{1,2}$ \cite{Haber1}. In order to obtain masses of the
neutral CP-even Higgs reasonably, we should include the radiative corrections from exotic fermions
and corresponding supersymmetric partners in the our model. Then, the mass square matrix for the
neutral CP-even Higgs in the basis  $(H_d^0,\;H_u^0)$ is written as
\begin{eqnarray}
&&{\cal M}^2_{even}=\left(\begin{array}{ll}M_{11}^2+\Delta_{11}&M_{12}^2+\Delta_{12}\\
M_{12}^2+\Delta_{12}&M_{22}^2+\Delta_{22}\end{array}\right)\;,
\label{M-CPE}
\end{eqnarray}
where
\begin{eqnarray}
&&M_{11}^2=m_{_{\rm Z}}^2\cos^2\beta+m_{_{A^0}}^2\sin^2\beta\;,
\nonumber\\
&&M_{12}^2=-(m_{_{\rm Z}}^2+m_{_{A^0}}^2)\sin\beta\cos\beta\;,
\nonumber\\
&&M_{22}^2=m_{_{\rm Z}}^2\sin^2\beta+m_{_{A^0}}^2\cos^2\beta\;,
\label{M-CPE1}
\end{eqnarray}
and $m_{_{A^0}}$ stands for the pseudo-scalar Higgs mass at tree level.
In this model the radiative corrections originate from the MSSM sector, exotic fermions and their scalar partners
respectively:
\begin{eqnarray}
&&\Delta_{11}=\Delta_{11}^{MSSM}+\Delta_{11}^{B}+\Delta_{11}^{L}\;,
\nonumber\\
&&\Delta_{12}=\Delta_{12}^{MSSM}+\Delta_{12}^{B}+\Delta_{12}^{L}\;,
\nonumber\\
&&\Delta_{22}=\Delta_{22}^{MSSM}+\Delta_{22}^{B}+\Delta_{22}^{L}\;.
\label{M-CPE2}
\end{eqnarray}
The concrete expressions for $\Delta_{11}^{MSSM}$, $\Delta_{12}^{MSSM}$, $\Delta_{22}^{MSSM}$
at two-loop level can be found in literatures \cite{2loopH,2loopH-1,2loopH-2,2loopH-3},
whereas the one-loop radiative corrections from the exotic quark
fields to $\Delta_{11}^{B}$, $\Delta_{12}^{B}$, $\Delta_{22}^{B}$ are formulated in Appendix \ref{app1}.
Considered that the VEVs of scalar components of $\hat{\Phi}_{_L}$ and $\hat{\varphi}_{_L}$ can induce the TeV
masses to the exotic leptons, we could choose sufficiently small exotic lepton Yukawa couplings and then  the radiative
corrections from exotic lepton fields for $\Delta_{11}^{L}$, $\Delta_{12}^{L}$, $\Delta_{22}^{L}$
can be ignored in our following numerical computations.

One of the most stringent constraints on the parameter space of the BLMSSM is that the mass square
matrix in Eq.~(\ref{M-CPE}) should produce an eigenvalue around $(125\;{\rm GeV})^2$
as mass square of the lightest neutral CP-even Higgs.
The current combination of the ATLAS and CMS data gives \cite{ATLAS,CMS,CMS-ATLAS}:
\begin{eqnarray}
&&m_{_{h^0}}=125.09\pm0.24\;{\rm GeV}\;,
\label{M-h0}
\end{eqnarray}
and this requirement restricts the parameter space of the BLMSSM strongly.
Besides the observed signals for the diphoton and $ZZ^*,\;WW^*,\;b\bar b$ channels of the 125 GeV Higgs  obtained by
the ATLAS and CMS collaborations are quantified by the ratios  \cite{Feng}
\begin{eqnarray}
&&R_{\gamma\gamma}={\Gamma_{_{NP}}(h^0\rightarrow gg)\Gamma_{_{NP}}(h^0\rightarrow\gamma\gamma)\over
\Gamma_{_{SM}}(h^0\rightarrow gg)\Gamma_{_{SM}}(h^0\rightarrow\gamma\gamma)}
\;,\nonumber\\
&&R_{VV^*}={\Gamma_{_{NP}}(h^0\rightarrow gg)\Gamma_{_{NP}}(h^0\rightarrow VV^*)\over
\Gamma_{_{SM}}(h^0\rightarrow gg)\Gamma_{_{SM}}(h^0\rightarrow VV^*)}
\;,\;\;(V=Z,\;W)\;.
\label{signal}
\end{eqnarray}
The weighted averages of the ratios are
\cite{Rrr-1,Rrr-2,Rrr-3,Rrr-4,Rrr-5,Rrr-6,Rrr-7,Rrr-8,Rrr-9,Rrr-10}:
\begin{eqnarray}
&&{\rm ATLAS+CMS}:\;\;R_{\gamma\gamma}=1.19\pm0.31\;,
\nonumber\\
&&{\rm ATLAS+CMS}:\;\;R_{VV^*}=0.86\pm0.16\;.
\label{signal-exp}
\end{eqnarray}
In the following numerical computations, we use the weighted averages of the ratios within $2\sigma$ tolerance to constrain the parameter space.

From Eq.~(\ref{CPevenB-mass}) and Eq.~(\ref{CPevenL-mass}), the masses of `brand new' neutral
Higgs satisfy tree-level relations
\begin{eqnarray}
&&m_{_{Z_B}}^2+m_{_{A_B^0}}^2=m_{_{h_B^0}}^2+m_{_{H_B^0}}^2\;,
\nonumber\\
&&m_{_{Z_B}}^2m_{_{A_B^0}}^2\cos^22\theta_{_B}=m_{_{h_B^0}}^2m_{_{H_B^0}}^2\;,
\nonumber\\
&&m_{_{Z_L}}^2+m_{_{A_L^0}}^2=m_{_{h_L^0}}^2+m_{_{H_L^0}}^2\;,
\nonumber\\
&&m_{_{Z_L}}^2m_{_{A_L^0}}^2\cos^22\theta_{_L}=m_{_{h_L^0}}^2m_{_{H_L^0}}^2\;.
\label{NeutralBLScalars}
\end{eqnarray}
When the radiative corrections do not modified those relations drastically, there
are several particularly interesting predictions:
\begin{eqnarray}
&&m_{_{h_B^0}}\le(m_{_{A_B^0}},m_{_{Z_B}})\le m_{_{H_B^0}}\;,\nonumber\\
&&m_{_{h_B^0}}\le\min(m_{_{A_B^0}},m_{_{Z_B}})|\cos2\theta_{_B}|\le m_{_{Z_B}}
\;,\nonumber\\
&&m_{_{h_L^0}}\le(m_{_{A_L^0}},m_{_{Z_L}})\le m_{_{H_L^0}}
\;,\nonumber\\
&&m_{_{h_L^0}}\le\min(m_{_{A_L^0}},m_{_{Z_L}})|\cos2\theta_{_L}|\le m_{_{Z_L}}\;.
\label{MassRelation}
\end{eqnarray}
It is not worth surprising because there are similar tree-level relations in the MSSM
which are modified drastically by the radiative corrections originating from large
Yukawa couplings of top and its superpartners.

Because of the Landau-Yang theorem~\cite{Landau,CNYang}, the 750 GeV resonance
with diphoton decay mode cannot be interpreted as the massive gauge bosons
$Z_{_B},\;Z_{_L}$ in this model. In addition, the 750 GeV resonance generally cannot
be interpreted as $H^0,\;A^0$ which are originated from $SU(2)$ doublets since we do not
find the resonance in the $WW$, $ZZ$, and $t\bar{t}$ channels. For the points mentioned above,
the potential candidate in the model considered here for the 750 GeV resonance
is possibly one of $h_{B,L}^0,\;H_{B,L}^0,\;A_{B,L}^0$. Nevertheless the leading order
contributions to $\Phi\rightarrow gg$ emerge at the 3-loop level, if we
took the 750 GeV resonance as one of $\Phi=h_L^0,\;H_L^0,\;A_L^0$. By the mass relations given
in Eq.~(40), we reasonably choose one of $\Phi=h_B^0,\;A_B^0$ to be
the 750 GeV resonance and $m_{_{Z_{B,L}}}\ge1$ TeV in accord with the
experimental constraint set by $Z^\prime$ searching at colliders~\cite{PDG}.


The 750 GeV scalar is produced mainly through the gluon fusion at the LHC.
In the supersymmetric extension of the SM,
the LO decay width for the process $\Phi\rightarrow gg$ ($\Phi=H^0,\;h_{_B}^0,\;H_{_B}^0$) is given as (see Refs.~\cite{Gamma1,Gamma1-1,Gamma1-2,Gamma1-3,Higgs-Hunter} and references therein)
\begin{eqnarray}
&&\Gamma_{_{NP}}(\Phi\rightarrow gg)={G_{_F}\alpha_s^2m_{_{\Phi}}^3\over64\sqrt{2}\pi^3}
\Big|\sum\limits_qg_{_{\Phi qq}}A_{1/2}(x_q)
+\sum\limits_{\tilde q}g_{_{\Phi\tilde{q}\tilde{q}}}{m_{_{\rm Z}}^2\over m_{_{\tilde q}}^2}A_{0}(x_{_{\tilde{q}}})\Big|^2\;,
\label{hgg}
\end{eqnarray}
with $x_a=m_{_{\Phi}}^2/(4m_a^2)$. In addition, $q=t,\;b,\;t_{_4},\;t_{_5},\;b_{_4},\;b_{_5}$
and $\tilde{q}=\tilde{t}_{_{1,2}},\;\tilde{b}_{_{1,2}},\;\tilde{\cal U}_i,\;\tilde{\cal D}_i\;(i=1,\;2,\;3,\;4)$.
The concrete expressions for $g_{_{\Phi tt}},\;g_{_{\Phi bb}},\;g_{_{\Phi\tilde{t}_i\tilde{t}_i}}
,\;g_{_{\Phi\tilde{b}_i\tilde{b}_i}},\;(i=1,\;2)$ can be found in the Refs. \cite{BL_h1,Gamma1-2}, and
the concrete expressions of $g_{_{\Phi t_{(i+3)}t_{(i+3)}}}$, $g_{_{\Phi b_{(i+3)}b_{(i+3)}}}$,
$g_{_{\Phi\tilde{\cal U}_i\tilde{\cal U}_i}}$, as well as $g_{_{\Phi\tilde{\cal D}_i\tilde{\cal D}_i}}$
are collected in Appendix \ref{PhiquarkCoupling}.

The form factors $A_{1/2}(x),\;A_0(x)$ in Eq.~(\ref{hgg}) are defined as
\begin{eqnarray}
&&A_{1/2}(x)=2\Big[x+(x-1)g(x)\Big]/x^2\;,\nonumber\\
&&A_0(x)=-(x-g(x))/x^2\;,
\label{loop-function1}
\end{eqnarray}
with
\begin{eqnarray}
&&g(x)=\left\{\begin{array}{l}\arcsin^2\sqrt{x},\;x\le1\\
-{1\over4}\Big[\ln{1+\sqrt{1-1/x}\over1-\sqrt{1-1/x}}-i\pi\Big]^2,\;x>1\;.\end{array}\right.
\label{g-function}
\end{eqnarray}

For the CP-odd scalar $\Phi=A^0,\;A_{_B}^0$, the decay width is written as
\begin{eqnarray}
&&\Gamma_{_{NP}}(\Phi\rightarrow gg)={G_{_F}\alpha_s^2m_{_{\Phi}}^3\over64\sqrt{2}\pi^3}
\Big|\sum\limits_qg_{_{\Phi qq}}A_{1/2}^\prime(x_q)\Big|^2\;,
\label{Agg}
\end{eqnarray}
with
\begin{eqnarray}
&&A_{1/2}^\prime(x)=2g(x)/x\;.
\label{loop-function3}
\end{eqnarray}

In the SM, the LO contributions to the diphoton decay of Higgs are derived from the one
loop diagrams containing virtual charged gauge boson $W^\pm$ or virtual top quarks.
Whereas in the BLMSSM, the exotic fermions $t_{_{4,5}},\;b_{_{4,5}},\;e_{_{4,5}}$
together with their supersymmetric partners contribute the corrections to the diphoton decay width
of CP-even neutral scalar at LO, the corresponding expression is written as
\begin{eqnarray}
&&\Gamma_{_{NP}}(\Phi\rightarrow\gamma\gamma)={G_{_F}\alpha^2m_{_{\Phi}}^3\over128\sqrt{2}\pi^3}
\Big|\sum\limits_fN_cQ_{_f}^2g_{_{\Phi ff}}A_{1/2}(x_{_f})+g_{_{\Phi WW}}A_1(x_{_{\rm W}})
\nonumber\\
&&\hspace{3.2cm}
+g_{_{\Phi H^+H^-}}{m_{_{\rm W}}^2\over m_{_{H^\pm}}^2}A_0(x_{_{H^\pm}})
+\sum\limits_{i=1}^2g_{_{\Phi\chi_i^+\chi_i^-}}{m_{_{\rm W}}\over m_{_{\chi_i}}}A_{1/2}(x_{_{\chi_i}})
\nonumber\\
&&\hspace{3.2cm}
+\sum\limits_{\tilde f}N_cQ_{_f}^2g_{_{\Phi\tilde{f}\tilde{f}}}{m_{_{\rm Z}}^2\over m_{_{\tilde f}}^2}
A_{0}(x_{_{\tilde{f}}})\Big|^2\;,
\label{hpp}
\end{eqnarray}
where $g_{_{h^0WW}}=\sin(\beta-\alpha),\;g_{_{H^0WW}}=\cos(\beta-\alpha)$, and
$g_{_{h_{_B}^0WW}}=g_{_{H_{_B}^0WW}}=0$, the loop function $A_1$ is
\begin{eqnarray}
&&A_1(x)=-\Big[2x^2+3x+3(2x-1)g(x)\Big]/x^2\;.
\label{loop-function2}
\end{eqnarray}
The concrete expressions for $g_{_{h^0(H^0)\chi_i^+\chi_i^-}},\;g_{_{h^0(H^0)H^+H^-}}$
and the couplings between the lightest neutral CP-even Higgs and exotic leptons/sleptons can also
be found in literature~\cite{BL_h1}. Furthermore one has $g_{_{h_{_B}^0\chi_i^+\chi_i^-}}=
g_{_{H_{_B}^0\chi_i^+\chi_i^-}}=0,\;g_{_{h_{_B}^0H^+H^-}}=g_{_{H_{_B}^0H^+H^-}}=0$.

Similarly the decays $\Phi\rightarrow Z\gamma\;(\Phi=h^0,\;H^0,\;h_{_B}^0,\;H_{_B}^0)$
are induced through loops involving massive charged particles which couple to
the scalar $\Phi$, those corresponding decay widths are formulated as
\begin{eqnarray}
&&\Gamma_{_{NP}}(\Phi\rightarrow Z\gamma)={G_{_F}\alpha^2m_{_{\Phi}}^3\over64\sqrt{2}s_{_{\rm W}}^2\pi^3}
\Big(1-{m_{_{\rm Z}}^2\over m_{_{\Phi}}^2}\Big)^2\Big|2\sum\limits_fN_cQ_{_f}{T_{_f}^{3L}-2Q_{_f}s_{_{\rm W}}^2
\over c_{_{\rm W}}}g_{_{\Phi ff}}A_{1/2}^h(x_{_f},y_{_f})
\nonumber\\
&&\hspace{3.2cm}
+g_{_{\Phi WW}}A_1^h(x_{_{\rm W}},y_{_{\rm W}})
+{2c_{_{\rm W}}^2-1\over2c_{_{\rm W}}^2}g_{_{\Phi H^+H^-}}{m_{_{\rm W}}^2\over m_{_{H^\pm}}^2}A_0^h(x_{_{H^\pm}},y_{_{H^\pm}})
\nonumber\\
&&\hspace{3.2cm}
+\sum\limits_{i=1}^2\sum\limits_{\alpha=L,R}{m_{_{\rm W}}\over m_{_{\chi_i}}}
g_{_{\Phi\chi_i^+\chi_i^-}}^\alpha g_{_{Z\chi_i^+\chi_i^-}}^\beta A_{1/2}^h(x_{_{\chi_i}},y_{_{\chi_i}})
\nonumber\\
&&\hspace{3.2cm}
+\sum\limits_{\tilde f}N_cQ_{_f}{T_{_f}^{3L}-Q_{_f}s_{_{\rm W}}^2
\over c_{_{\rm W}}}g_{_{\Phi\tilde{f}\tilde{f}}}{m_{_{\rm Z}}^2\over m_{_{\tilde f}}^2}
A_{0}^h(x_{_{\tilde{f}}},y_{_{\tilde{f}}})\Big|^2\;,
\label{hZphoton}
\end{eqnarray}
where $y_{_i}=m_{_{\rm Z}}^2/(4m_{_i}^2)$, and $T_{_f}^{3L}=\pm1/2$ denotes the 3rd component
of weak isospin of corresponding matter field. For convenience the form factors are written as
\begin{eqnarray}
&&A_0^h(x,y)=I_1(x,y)\;,\nonumber\\
&&A_{1/2}^h(x,y)=I_1(x,y)-I_2(x,y)\;,\nonumber\\
&&A_1^h(x,y)=c_{_{\rm W}}\Big\{4(3-{s_{_{\rm W}}^2\over c_{_{\rm W}}^2})I_2(x,y)
+\Big[(1+2x){s_{_{\rm W}}^2\over c_{_{\rm W}}^2}-(5+2x)\Big]I_1(x,y)\Big\}\;,
\label{hZphoton-FormFactor}
\end{eqnarray}
with
\begin{eqnarray}
&&I_1(x,y)=-{1\over2(x-y)}+{g(x)-g(y)\over2(x-y)^2}+{y(f(x)-f(y))\over2(x-y)^2}\;,\nonumber\\
&&I_2(x,y)={g(x)-g(y)\over2(x-y)}\;,\nonumber\\
&&f(x)=\left\{\begin{array}{l}\sqrt{x-1}\arcsin^2\sqrt{1/x},\;x\leq1\\
{\sqrt{1-x}\over2}\Big[\ln{1+\sqrt{1-x}\over1-\sqrt{1-x}}-i\pi\Big]^2,\;x>1\;.\end{array}\right.
\label{hZphoton-Functions}
\end{eqnarray}
Generally the signature of this decay mode
is drowned in the huge background from $q\bar{q}\rightarrow Z\gamma$~\cite{Higgs-Hunter,JFGunion}
and $gg\rightarrow Z\gamma$~\cite{JJvanderBij}.

The LO diphoton decay width of the CP-odd neutral scalar $\Phi=A^0,\;A_{_B}^0$ ($\Phi\rightarrow\gamma\gamma$) is formulated as
\begin{eqnarray}
&&\Gamma_{_{NP}}(\Phi\rightarrow\gamma\gamma)={G_{_F}\alpha^2m_{_{\Phi}}^3\over128\sqrt{2}\pi^3}
\Big|\sum\limits_fN_cQ_{_f}^2g_{_{\Phi ff}}A_{1/2}^\prime(x_f)
+\sum\limits_{i=1}^2g_{_{\Phi\chi_i^+\chi_i^-}}{m_{_{\rm W}}\over m_{_{\chi_i}}}A_{1/2}^\prime(x_{_{\chi_i}})\Big|^2\;.
\label{A0pp}
\end{eqnarray}
In a similar way, we can write down the decay widths for $\Phi\rightarrow Z\gamma$
for $\Phi=A^0,\;A_{_B}^0$ as
\begin{eqnarray}
&&\Gamma_{_{NP}}(\Phi\rightarrow Z\gamma)={G_{_F}\alpha^2m_{_{\Phi}}^3\over64\sqrt{2}s_{_{\rm W}}^2\pi^3}
\Big(1-{m_{_{\rm Z}}^2\over m_{_{\Phi}}^2}\Big)^2
\Big|2\sum\limits_fN_cQ_{_f}{T_{_f}^{3L}-2Q_{_f}s_{_{\rm W}}^2
\over c_{_{\rm W}}}g_{_{\Phi ff}}A_{1/2}^a(x_{_f},y_{_f})
\nonumber\\
&&\hspace{3.2cm}
+\sum\limits_{i=1}^2\sum\limits_{\alpha=L,R}{m_{_{\rm W}}\over m_{_{\chi_i}}}
g_{_{\Phi\chi_i^+\chi_i^-}}^\alpha g_{_{Z\chi_i^+\chi_i^-}}^\beta A_{1/2}^a(x_{_{\chi_i}},y_{_{\chi_i}})\Big|^2\;,
\label{A0Zphoton}
\end{eqnarray}
with $A_{1/2}^a(x,y)=I_2(x,y)$.

The neutral scalar with mass around $750\;{\rm GeV}$ would decay through the modes
$\Phi\rightarrow ZZ,\;\Phi\rightarrow WW,\;\Phi\rightarrow \bar{t}t$,
where $Z/W$ denote the on-shell neutral/charged electroweak gauge bosons and the corresponding widths are:
~\cite{Higgs-Hunter,HtoVV-SUSY,HtoVV-SUSY-1,HtoVV-SUSY-2}
\begin{eqnarray}
&&\Gamma_{_{NP}}(\Phi\rightarrow \bar{t}t)={3G_{_F}m_{_t}^2\over4\sqrt{2}\pi}|g_{_{\Phi tt}}|^2
m_{_{\Phi}}\beta_{_t}^{p(\Phi)}\Big[1+{4\alpha_{_S}\over3\pi}\Delta_{_\Phi}(\beta_{_t})\Big],\;\nonumber\\
&&\Gamma_{_{NP}}(\Phi\rightarrow WW)={G_{_F}\over8\sqrt{2}\pi}m_{_{\Phi}}^3|g_{_{\Phi WW}}|^2
\sqrt{1-x_{_{\rm W}}}(1-x_{_{\rm W}}+{3\over4}x_{_{\rm W}}^2)
\Big[1+0.175{G_{_F}m_{_\Phi}^2\over\sqrt{2}\pi^2}\Big],\;\nonumber\\
&&\Gamma_{_{NP}}(\Phi\rightarrow ZZ)={G_{_F}\over16\sqrt{2}\pi}m_{_{\Phi}}^3|g_{_{\Phi ZZ}}|^2
\sqrt{1-x_{_{\rm Z}}}(1-x_{_{\rm Z}}+{3\over4}x_{_{\rm Z}}^2)
\Big[1+0.175{G_{_F}m_{_\Phi}^2\over\sqrt{2}\pi^2}\Big]\;,
\label{Phi-WWZZ}
\end{eqnarray}
with $g_{_{h^0ZZ}}=g_{_{h^0WW}}$, and $x_{_V}=4m_{_V}^2/m_{_\Phi}^2\;(V=W,\;Z)$.
Meanwhile the radiative corrections
\begin{eqnarray}
&&\Delta_{_{\Phi}}(\beta_{_t})=\left\{\begin{array}{l}
{1\over\beta_{_t}}A(\beta_{_t})+{1\over16\beta_{_t}^3}
(3+34\beta_{_t}^2-13\beta_{_t}^4)\ln{1+\beta_{_t}\over1-\beta_{_t}}
+{3\over8\beta_{_t}^2}(7\beta_{_t}^2-1),\Phi=H^0,h_{_B}^0,H_{_B}^0\quad\\
{1\over\beta_{_t}}A(\beta_{_t})+{1\over16\beta_{_t}^3}
(19+2\beta_{_t}^2+3\beta_{_t}^4)\ln{1+\beta_{_t}\over1-\beta_{_t}}
+{3\over8\beta_{_t}^2}(7-\beta_{_t}^2),\Phi=A^0,A_{_B}^0\end{array}\right.
\label{radiative-corrections}
\end{eqnarray}
with $\beta_{_t}^2=1-4m_{_t}^2/m_{_\Phi}^2,\;p(H^0)=p(h_{_B}^0)=p(H_{_B}^0)=3,\;p(A^0)=p(A_{_B}^0)=1$, and
\begin{eqnarray}
&&A(\beta_{_t})=(1+\beta_{_t}^2)\Big[4Li_2({1-\beta_{_t}\over1+\beta_{_t}})+2Li_2({\beta_{_t}-1\over1+\beta_{_t}})
+3\ln{1-\beta_{_t}\over1+\beta_{_t}}\ln{2\over1+\beta_{_t}}
\nonumber\\
&&\hspace{1.5cm}
+2\ln{1-\beta_{_t}\over1+\beta_{_t}}\ln \beta_{_t}\Big]
-4\beta_{_t}\ln{4\beta_{_t}^{4/3}\over1-\beta_{_t}^2}\;.
\label{radiative-funtions}
\end{eqnarray}
The loop induced couplings $g_{_{\Phi tt}},\;g_{_{\Phi ZZ}},\;g_{_{\Phi WW}}\;(\Phi=h_{_B}^0,\;H_{_B}^0,\;A_{_B}^0)$
are given in Appendix~\ref{LoopInduceCoupling}.

Considering the fact that no 750 GeV diphoton excess was observed at 8 TeV run of LHC ~\cite{LHC8,LHC8-1} but an excess shows up
at 13 TeV~\cite{LHC13,LHC13-1}, we should determine that the heavy scalar most likely is produced via gluon fusion at 13 TeV.
Therefore, the observed signals for the scalar diphoton excess at the LHC can be quantified as
\begin{eqnarray}
&&\mu_{13\rm{TeV}}^{\Phi}= \sigma(gg\rightarrow \Phi)\; {{\rm{BR}}}(\Phi\rightarrow \gamma\gamma)\;\nonumber\\
&&\hspace{1.25cm}
= \sigma(gg\rightarrow \Phi)\; \Gamma_{_{NP}}(\Phi\rightarrow \gamma\gamma)/\Gamma^{\rm{tot}}_\Phi\;.
\label{750-signals1}
\end{eqnarray}
The total decay width of $\Phi$ is
\begin{eqnarray}
&&\Gamma^{\rm{tot}}_\Phi= \Gamma_{_{NP}}(\Phi\rightarrow gg)
+ \Gamma_{_{NP}}(\Phi\rightarrow \gamma\gamma)
+ \Gamma_{_{NP}}(\Phi\rightarrow Z\gamma)\nonumber\\
&&\hspace{1.25cm}
+ \Gamma_{_{NP}}(\Phi\rightarrow ZZ)
+ \Gamma_{_{NP}}(\Phi\rightarrow WW)
+ \Gamma_{_{NP}}(\Phi\rightarrow \bar{t}t)+\Gamma_{_{NP}}^{other},
\label{750-signals2}
\end{eqnarray}
where $\Gamma_{_{NP}}^{other}$ denotes the width for other decay modes of $\Phi$.
Due that $\sigma(gg\rightarrow \Phi)\propto \Gamma(\Phi\rightarrow gg)$, we could have
\begin{eqnarray}
\sigma(gg\rightarrow \Phi)={ \Gamma_{_{NP}}(\Phi\rightarrow gg)\over \Gamma_{_{NP}}(h^0 \rightarrow gg)} \sigma(gg\rightarrow h^0)|_{m_{_{h^0}}\simeq750\;{\rm GeV}}\:,
\end{eqnarray}
where $\sigma(gg\rightarrow h^0)\approx0.85 \times10^3 \;\rm{fb}$~\cite{sigma-ggh,sigma-ggh1}. The combined value of 8 and 13 TeV measurements roughly is~\cite{NSoneS2}
\begin{eqnarray}
&&\mu_{13\rm{TeV}}^{exp}= (4.4\:\pm\:1.1)\;\rm{fb}.
\label{750-exp}
\end{eqnarray}
In the following numerical calculation, we will take into account the combined experimental value at $3\sigma$ as a simply guideline.

\section{Numerical analyses\label{sec4}}
\indent\indent
To proceed our numerical discussion, we choose relevant parameters of the SM as~\cite{PDG}
\begin{eqnarray}
&&\alpha_s(m_{_{\rm Z}})=0.118\;,\;\;\alpha(m_{_{\rm Z}})=1/128\;,
\;\;s_{_{\rm W}}^2(m_{_{\rm Z}})=0.23\;,
\nonumber\\
&&m_t=174.2\;{\rm GeV}\;,\;\;m_b=4.2\;{\rm GeV}\;,\;\;
m_{_{\rm W}}=80.4\;{\rm GeV}\;.
\label{PDG-SM}
\end{eqnarray}
As aforementioned, the most stringent constraint on the parameter space is that the
$2\times2$ mass square matrix in Eq.~(\ref{M-CPE}) whose lightest eigenvector must be of
a mass $m_{_{h^0}}\simeq125.09\pm0.24\;{\rm GeV}$. In order to obtain the final results satisfying this constraint,
we require the tree level mass of CP-odd Higgs $m_{_{A^0}}$ to be
\begin{eqnarray}
&&m_{_{A^0}}^2={m_{_{h^0}}^2(m_{_{\rm z}}^2-m_{_{h^0}}^2+\Delta_{_{11}}+\Delta_{_{22}})-m_{_{\rm z}}^2
\Delta_{_A}+\Delta_{_{12}}^2-\Delta_{_{11}}\Delta_{_{22}}\over -m_{_{h^0}}^2+m_{_{\rm z}}^2\cos^22\beta
+\Delta_{_B}}\;,
\label{Higgs-mass1}
\end{eqnarray}
where
\begin{eqnarray}
&&\Delta_{_A}=\sin^2\beta\Delta_{_{11}}+\cos^2\beta\Delta_{_{22}}+\sin2\beta \Delta_{_{12}}
\;,\nonumber\\
&&\Delta_{_B}=\cos^2\beta\Delta_{_{11}}+\sin^2\beta\Delta_{_{22}}+\sin2\beta \Delta_{_{12}}\;.
\label{Higgs-mass2}
\end{eqnarray}

In order to avoid Landau singularities of $g_{_{B,L}}$ below the Planck scale,
we choose $B_{_4}=L_{_4}=0,\;g_{_B}(\Lambda_{_{NP}})=0.35,\;g_{_L}(\Lambda_{_{NP}})=0.2$
with $\Lambda_{_{NP}}=3$ TeV. Meanwhile we assume $m_{_{Z_B}}=m_{_{Z_L}}=1$ TeV
to coincide with experimental data of searching additional neutral gauge bosons in colliders~\cite{PDG}.
As discussed above, the plausible candidates for the 750 GeV resonance are $h_{_B}^0$
and $A_{_B}^0$ in this model. Since there is no correction from exotic leptons and
their superpartners to the diphoton channels $h_{_B}^0\rightarrow2\gamma$,
$A_{_B}^0\rightarrow2\gamma$ at leading order, moreover the corrections from exotic leptons and
their superpartners to $h^0\rightarrow2\gamma$ are negligible if those particle masses are of order TeVs.
In view of this, we could choose $\tan\beta_{_L}=2$,
$\lambda_{_L}=\lambda_{_E}=\lambda_{_N}=0.5$,
$m_{_{\tilde{L}_{4,5}}}=m_{_{\tilde{\nu}_{4,5}}}=m_{_{\tilde{E}_{4,5}}}=3\;{\rm TeV}$, 
$A_{{\nu_{4,5}}}=A_{{e_{4,5}}}=500\;{\rm GeV}$
in our numerical analyses. In order to predict the mass of $h^0$ falling in the range
$124\;{\rm GeV}\le m_{_{h^0}}\le 126\;{\rm GeV}$, we take
$m_{_{\tilde{Q}_3}}=1\;{\rm TeV}\;,\; m_{_{\tilde{U}_3}}=m_{_{\tilde{D}_3}}=2\;{\rm TeV}$,
$A_{_t}=2.1\;{\rm TeV}$, $A_{_b}=-1\;{\rm TeV}$,
$Y_{_{d_4}}=Y_{_{d_5}}=0.7\;Y_b$, and $\tan\beta=1.5$ unless a particular specification being made.

If we interpret the 750 GeV resonance as the CP-even scalar $h_{_B}^0$,
we find that the signal
$\mu_{13\rm{TeV}}^{h_{_B}^0}\le \mathcal{O}(10^{-1}\;{\rm fb})$
through scanning the parameter space of the model, because there is
a cancellation between corrections from exotic quarks charged $2/3$
and that charged $-1/3$. However the cancelation
does not appear as we interpret the 750 GeV resonance as the CP-odd scalar $A_{_B}^0$
with a mass around 750 GeV which can account for
the signal on  diphoton excess at $750\;{\rm GeV}$ observed by the ATLAS
and CMS collaborations simultaneously. Thus we choose the CP-odd scalar
$A_{_B}^0$ as the heavy boson and keep $m_{_{A_{_B}^0}}= 750\;{\rm GeV}$ in the following.

In CP-conserving circumstances the decay channels $A_{_B}^0\rightarrow\gamma\gamma,\;gg$
are not affected by those parameters originating from scalar quark sectors
at leading order, we take the parameters of corresponding squarks sector as
\begin{eqnarray}
&&m_{_{\tilde{Q}_4}}=m_{_{\tilde{U}_4}}=m_{_{\tilde{D}_4}}=m_{_{\tilde{Q}_5}}=m_{_{\tilde{U}_5}}
=m_{_{\tilde{D}_5}}=3\;{\rm TeV}\;,\nonumber\\
&&A_{_{u_4}}=A_{_{d_4}}=A_{_{u_5}}=A_{_{d_5}}=100\;{\rm GeV}\;,\nonumber\\
&&A_{_{BQ}}=A_{_{BU}}=A_{_{BD}}=1\;{\rm TeV}\;.
\label{assumption1}
\end{eqnarray}

Under our above assumptions on parameter space, we always take
\begin{eqnarray}
&&m_2=700\;{\rm GeV}\;,\;\;\mu_{_B}=500\;{\rm GeV}\;,\;\;\mu=-800\;{\rm GeV}\;,
\label{assumption4}
\end{eqnarray}
since those parameters affect our theoretical evaluations mildly.
Then, the free parameters affecting strongly our numerical results are
\begin{eqnarray}
\lambda_{_Q} ,\;\;\lambda_{_U} ,\;\;\lambda_{_D} ,\;\; \tan \beta ,\;\;
\tan \beta_{_B} ,\;\; Y_{_{u_4}} ,\;\; Y_{_{u_5}}\;.
\end{eqnarray}

\begin{figure}
\setlength{\unitlength}{1mm}
\centering
\includegraphics[width=3.1in]{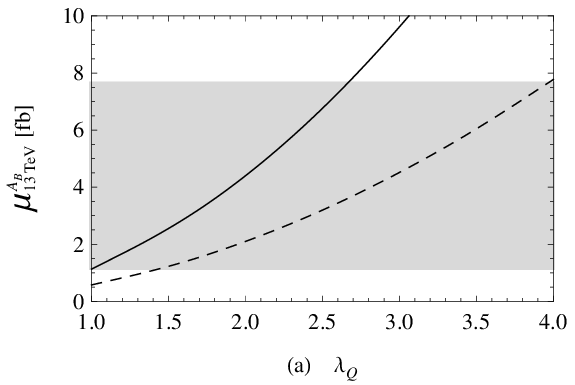}%
\vspace{0.5cm}
\includegraphics[width=3.1in]{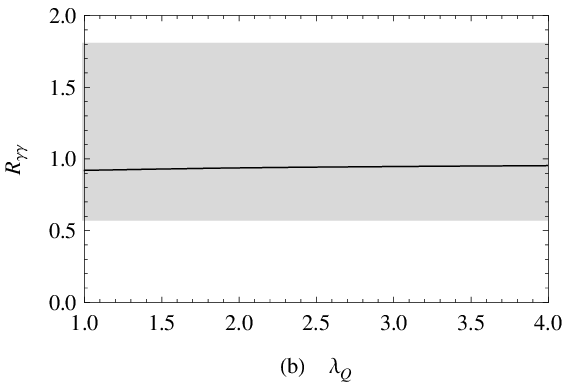}
\vspace{0cm}
\caption[]{As $Y_{_{u_4}}=0.2\; Y_t,\;Y_{_{u_5}}=0.4\; Y_t,\;\tan \beta=1.5$, and $\tan \beta_{_B}=3$,
(a) $\mu_{13\rm{TeV}}^{A_{_B}}\,[\rm{fb}]$ (solid line for $\lambda_{_U}=\lambda_{_D}=0.3$ and dashed line for $\lambda_{_U}=\lambda_{_D}=0.4$) varies with the parameter $\lambda_{_Q}$
where gray area denotes the experimental permission at $3\sigma$ deviations in Eq.~(\ref{750-exp}), (b)
$R_{\gamma\gamma}$ (for $\lambda_{_U}=\lambda_{_D}=0.3$) varies with the parameter $\lambda_{_Q}$ where gray
area denotes the experimental permission at $2\sigma$ deviations in
Eq.~(\ref{signal-exp}), respectively.}
\label{fig1}
\end{figure}

Taking $Y_{_{u_4}}=0.2\; Y_t,\;Y_{_{u_5}}=0.4\; Y_t,\;
\tan \beta=1.5$, and $\tan \beta_{_B}=3$,
we plot the signal $\mu_{13\rm{TeV}}^{A_{_B}}\,[\rm{fb}]$ (solid line for $\lambda_{_U}=\lambda_{_D}=0.3$ and dashed line for $\lambda_{_U}=\lambda_{_D}=0.4$) versus
parameter $\lambda_{_Q}$ in Fig.~\ref{fig1}(a), where gray area
denotes the experimental permission at $3\sigma$ deviations shown in Eq.~(\ref{750-exp}).
The numerical result indicates that the signal
$\mu_{13\rm{TeV}}^{A_{_B}}$ is consistent with the experimental data as
$1\le\lambda_{_Q}\le2.7$ for $\lambda_{_U}=\lambda_{_D}=0.3$ and
$1.4\le\lambda_{_Q}\le4$ for $\lambda_{_U}=\lambda_{_D}=0.4$. We can see that the signal
$\mu_{13\rm{TeV}}^{A_{_B}}$ turns stronger along with increasing of
$\lambda_{_Q}$ for the couplings in Eq.~(\ref{g-coupling5}) are proportional to $\lambda_{_Q}$.
On the contrary, the signal $\mu_{13\rm{TeV}}^{A_{_B}}$ turns smaller along with increasing of
$\lambda_{_{U,D}}$ since the mass of the lightest vector-like quark charged $2/3$ is
proportional to $\lambda_{_{U}}$, and that of the lightest vector-like quark charged $-1/3$
is proportional to $\lambda_{_{D}}$, respectively. In Fig.~\ref{fig1}(b), we show the signal
strength of the 125 GeV Higgs $R_{\gamma\gamma}$ varying with the
parameter $\lambda_{_Q}$ for $\lambda_{_U}=\lambda_{_D}=0.3$, where gray area denotes the experimental
permission at $2\sigma$ deviations in Eq.~(\ref{signal-exp}). We can see that the signal strength
$R_{\gamma\gamma}$ is gentle with increasing of $\lambda_{_Q}$.  The results indicate that
the signal strength $R_{\gamma\gamma}$ is consistent with the experimental
data. Similarly the  signal strength $R_{VV^*}$ can also fit the
experimental data in Eq.~(\ref{signal-exp}). The numerical results
implicate that the signal strength $R_{\gamma\gamma}$ also depends on the parameters
$\lambda_{_U}$ and $\lambda_{_D}$ mildly, actually the theoretical evaluations
on $R_{\gamma\gamma}$ varying with the parameter $\lambda_{_Q}$ for $\lambda_{_U}=\lambda_{_D}=0.4$
almost overlap with that for $\lambda_{_U}=\lambda_{_D}=0.3$.

\begin{figure}
\setlength{\unitlength}{1mm}
\centering
\includegraphics[width=3.1in]{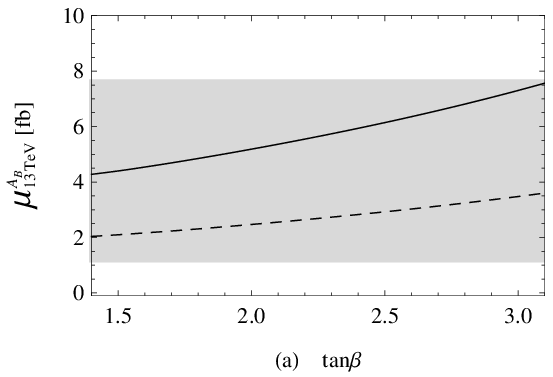}%
\vspace{0.5cm}
\includegraphics[width=3.1in]{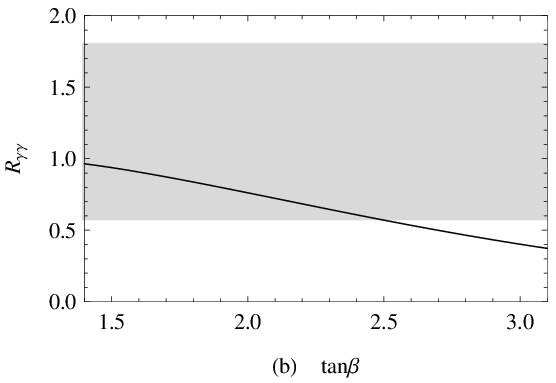}
\vspace{0cm} \caption[]{As $Y_{_{u_4}}=0.2\; Y_t,\;Y_{_{u_5}}=0.4\; Y_t,\;
\lambda_{_Q}=2$, and $\tan \beta_{_B}=3$, (a)
$\mu_{13\rm{TeV}}^{A_{_B}}$ (solid line for $\lambda_{_U}=\lambda_{_D}=0.3$ and dashed line for $\lambda_{_U}=\lambda_{_D}=0.4$) varies with the parameter $\tan \beta$
where gray area denotes the experimental permission at $3\sigma$ deviations in
Eq.~(\ref{750-exp}), and (b) $R_{\gamma\gamma}$ (for $\lambda_{_U}=\lambda_{_D}=0.3$) varies with the
parameter $\tan \beta$ where gray area denotes the experimental permission
at $2\sigma$ deviations in Eq.~(\ref{signal-exp}), respectively.}
\label{fig2}
\end{figure}

In addition, the ATLAS and CMS collaborations  showed that in Run I stage
no significant excesses were observed in the channels of 750 GeV  Higgs
decaying into $ZZ$ \cite{LHC-ZZ}, $WW$ \cite{LHC-WW,LHC-WW1} and $Z\gamma$ \cite{LHC-Zr}. As generally believed,
gluon fusion is responsible for the production of the Higgs boson which later may decay
into those final states, thus the data of LHC at 8 TeV
set upper bounds on the ratios as \cite{NS1}
\begin{eqnarray}
{\Gamma(\Phi\rightarrow Z\gamma)\over \Gamma(\Phi\rightarrow
\gamma\gamma)}<2\;,\;\;\;
{\Gamma(\Phi\rightarrow ZZ)\over \Gamma(\Phi\rightarrow
\gamma\gamma)}<6\;,\;\;\;{\Gamma(\Phi\rightarrow WW)\over \Gamma(\Phi\rightarrow \gamma\gamma)}<20\;.
\label{LHC-ZW}
\end{eqnarray}
In the chosen parameter space of the BLMSSM model,  $A_{_B}^0\rightarrow ZZ(WW)$ appears
at one-loop level and we have obtained the relevant ratios as
\begin{eqnarray}
{\Gamma(A_{_B}^0\rightarrow Z\gamma)\over \Gamma(A_{_B}^0\rightarrow \gamma\gamma)}
\sim \mathcal{O}(10^{-1}),\quad
{\Gamma(A_{_B}^0\rightarrow ZZ)\over \Gamma(A_{_B}^0\rightarrow \gamma\gamma)}
\sim \mathcal{O}(10^{-1}),\quad
{\Gamma(A_{_B}^0\rightarrow WW)\over \Gamma(A_{_B}^0\rightarrow \gamma\gamma)}\sim \mathcal{O}(1),
\end{eqnarray}
which confirm the bounds presented in Eq.~(\ref{LHC-ZW}).
In this model, the decay mode $\Gamma_{_{NP}}(A_{_B}^0\rightarrow t\bar{t})$
can only occur via two-loop diagrams, so its rate is  smaller than the width of diphoton channel.
Since, as generally expected, the 750 GeV resonance is produced via gluon fusion,
there is a large probability it would decay into two gluons which turn into di-jet.
In this work, the numerical result indicates that
${\Gamma(A_{_B}^0\rightarrow gg)/ \Gamma(A_{_B}^0\rightarrow \gamma\gamma)}
\sim \mathcal{O}(10^{2})<1300$, which accommodates the di-jet research at Run I \cite{NS1,LHC-gg,LHC-gg1}.

Besides the parameter $\lambda_{_Q}$, the parameter $\tan \beta$
existing in the MSSM also affects our numerical evaluations strongly.
Choosing $Y_{_{u_4}}=0.2\; Y_t,\;Y_{_{u_5}}=0.4\; Y_t,\;
\lambda_{_Q}=2$, and $\tan \beta_{_B}=3$, we depict in Fig.~\ref{fig2}(a)
the signal $\mu_{13\rm{TeV}}^{A_{_B}}$ (solid line for $\lambda_{_U}=\lambda_{_D}=0.3$ and dashed line for $\lambda_{_U}=\lambda_{_D}=0.4$)
versus $\tan \beta$ where gray area denotes the experimental permission
at $3\sigma$ deviations in Eq.~(\ref{750-exp}), and Fig.~\ref{fig2}(b) the signal strength
$R_{\gamma\gamma}$ (for $\lambda_{_U}=\lambda_{_D}=0.3$) versus  $\tan \beta$ where gray area
denotes the experimental permission at $2\sigma$ deviations in
Eq.~(\ref{signal-exp}), respectively.  Fig.~\ref{fig2}(a) shows
that the signal $\mu_{13\rm{TeV}}^{A_{_B}}$ turns large as
$\tan \beta$ increasing. When the parameter $\tan \beta >
3.1$ as $\lambda_{_U}=\lambda_{_D}=0.3$, the signal $\mu_{13\rm{TeV}}^{A_{_B}}$ exceeds the upper bound.
For $\lambda_{_U}=\lambda_{_D}=0.4$, the signal $\mu_{13\rm{TeV}}^{A_{_B}}$ is coincide with the experimental
data at $3\sigma$ deviations.
With increasing of $\tan \beta$, the signal strength $R_{\gamma\gamma}$ decreases.
As $\tan \beta> 2.5$ and $\lambda_{_U}=\lambda_{_D}=0.3$, we cannot account for
the experimental results for the signal
strength of the 125 GeV Higgs $R_{\gamma\gamma}$, showed in Fig.~\ref{fig2}(b).
In other words, the simultaneous interpretation of experimental data on the decays of the
heavy scalar with 750 GeV and the lightest Higgs with 125 GeV similarly favors relatively small value of $\tan\beta$
under our assumptions on the parameter space.

\begin{figure}
\setlength{\unitlength}{1mm}
\centering
\includegraphics[width=3.1in]{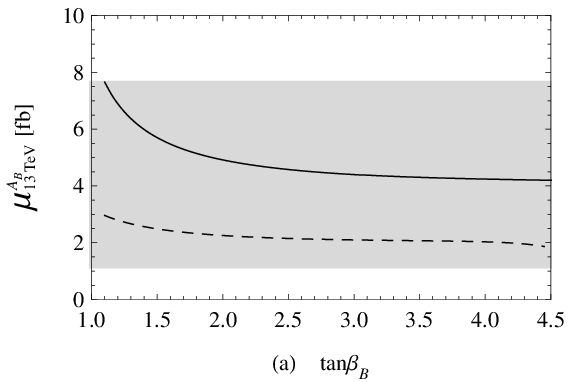}%
\vspace{0.5cm}
\includegraphics[width=3.1in]{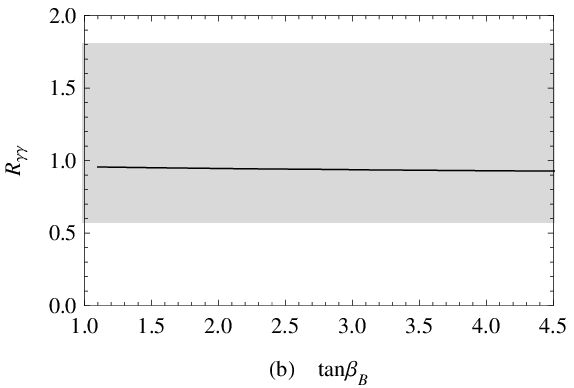}
\vspace{0cm} \caption[]{As $Y_{_{u_4}}=0.2\; Y_t,\;Y_{_{u_5}}=0.4\; Y_t,\;
\tan \beta=1.5$,
and $\lambda_{_Q}=2$, (a) $\mu_{13\rm{TeV}}^{A_{_B}}$ (solid line for $\lambda_{_U}=\lambda_{_D}=0.3$ and dashed line for $\lambda_{_U}=\lambda_{_D}=0.4$) varies with
the parameter $\tan \beta_{_B}$ where gray area denotes the
experimental permission at $3\sigma$ deviations in Eq.~(\ref{750-exp}), and (b)
$R_{\gamma\gamma}$ (for $\lambda_{_U}=\lambda_{_D}=0.3$) varies with the parameter $\tan \beta_{_B}$ where
gray area denotes the experimental permission at $2\sigma$ deviations in
Eq.~(\ref{signal-exp}), respectively.} \label{fig3}
\end{figure}

\begin{figure}
\setlength{\unitlength}{1mm}
\centering
\includegraphics[width=3.1in]{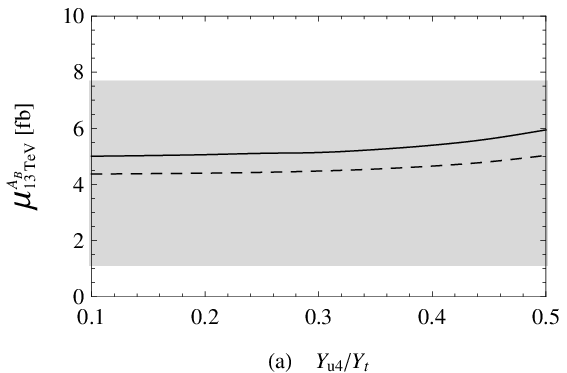}%
\vspace{0.5cm}
\includegraphics[width=3.1in]{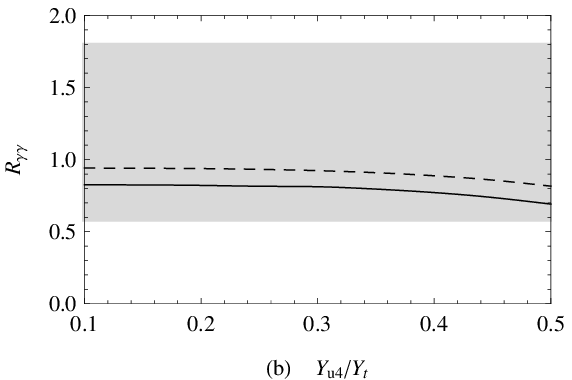}
\vspace{0cm}
\caption[]{As $\lambda_{_U}=\lambda_{_D}=0.3$, $\lambda_{_Q}=2,\;\tan \beta=1.5$, and $\tan \beta_{_B}=3$,
(a) $\mu_{13\rm{TeV}}^{A_{_B}}$ varies with the parameter $Y_{_{u_4}}$ where gray area
denotes the experimental permission at $3\sigma$ deviations in Eq.~(\ref{750-exp}), and (b) $R_{\gamma\gamma}$
varies with the parameter $Y_{_{u_4}}$ where gray area denotes
the experimental permission at $2\sigma$ deviations in Eq.~(\ref{signal-exp}), respectively.
Here, the dashed line stands for $Y_{_{u_5}}=0.4\; Y_t$, the solid line stands for $Y_{_{u_5}}=0.6\; Y_t$. }
\label{fig4}
\end{figure}

Additional the parameter $\tan\beta_{_B}$ in this model also affects our
numerical results strongly. In Fig.~\ref{fig3},
we investigate (a) the signal strength $\mu_{13\rm{TeV}}^{A_{_B}}$ (solid line for $\lambda_{_U}=\lambda_{_D}=0.3$ and dashed line for $\lambda_{_U}=\lambda_{_D}=0.4$) and
(b) the signal strength $R_{\gamma\gamma}$ (for $\lambda_{_U}=\lambda_{_D}=0.3$) varying with the parameter $\tan \beta_{_B}$,
where $Y_{_{u_4}}=0.2\; Y_t,\;Y_{_{u_5}}=0.4\; Y_t,\;\tan \beta=1.5$,
and $\lambda_{_Q}=2$. It is seen that the signal strength $\mu_{13\rm{TeV}}^{A_{_B}}$
decreases steeply as $\tan \beta_{_B}<2$, and decreases mildly as $\tan \beta_{_B}<3$.
As for the signal strength $R_{\gamma\gamma}$ varies with $\tan \beta_{_B}$ slowly.

At the last, we investigate the Yukawa couplings of the fourth and fifth generation
up-type quark $Y_{_{u_{4,5}}}$ in Fig.~\ref{fig4}. Taking $\lambda_{_U}=\lambda_{_D}=0.3$,
$\lambda_{_Q}=2,\;\tan \beta=1.5$ and $\tan \beta_{_B}=3$, we
plot the signal strength $\mu_{13\rm{TeV}}^{A_{_B}}$ versus
$Y_{_{u_4}}$ in (a) and the signal strength $R_{\gamma\gamma}$ vs
$Y_{_{u_4}}$ in (b) of Fig.~\ref{fig4}, where the dashed line stands for $Y_{_{u_5}}=0.4\; Y_t$
and the solid line stands for $Y_{_{u_5}}=0.6\; Y_t$, respectively.
With increasing of $Y_{_{u_4}}$, the signal strength $\mu_{13\rm{TeV}}^{A_{_B}}$ turns stronger,
on the other hand the signal strength $R_{\gamma\gamma}$ turns small.
In other words the large Yukawa couplings $Y_{_{u_5}}$ affects our numerical evaluations on
the signal strength $\mu_{13\rm{TeV}}^{A_{_B}}$ and the signal strength $R_{\gamma\gamma}$
simultaneously.

\section{Summary\label{sec5}}
\indent\indent
The discovery of 750 GeV boson at the diphoton channel is very inspiring because it
may be a signal for new physics BSM. People are excited and tempted to try various models in hand
to investigate the case and see if the model with a ceratin parameter range can give a reasonable
interpretation. We argue that an extension of the supersymmetric model with
gauged baryon and lepton numbers might be able to account for the
experimental data on 750 GeV diphoton excess reported by ATLAS and
CMS recently based on its success in earlier phenomenological studies.

Indeed, even though the 750 GeV boson is observed in the diphoton channel  as a resonance,
there are still many puzzles about its eccentric behaviors are not well understood yet. The first
challenge is its unusually large width about 45 GeV reported by ATLAS while CMS shows that it could be small. And it is also reported that this resonance is not
seen at the $WW,\; ZZ,\;$ and $t\bar t$ channels. It implies that it has some decay channels which are
not experimental observed yet, secondly, its coupling to the regular SM particles must be very suppressed, or
just as the diphoton channel the effective coupling to SM particles is realized via loops inside which only heavy
BSM particles exist.

In this BLMSSM, because the scalar $h_{_L}^0,
H_{_L}^0,A_{_L}^0$ do not have couplings to the exotic quarks at tree
level, they can be ruled out for being a candidate of the scalar
particle of $m_{_\Phi}=750$ GeV observed at the diphoton channel.
The contribution of $H^0, A^0$ and $h_{_B}^0$ to the diphoton decay widths
is too small to be responsible for the diphoton excess even though their mass were 750 GeV.
By contrary,
adopting an assumption on the
relevant parameter space, the CP-odd scalar $A_{_B}^0$ with
$750\;{\rm GeV}$ mass  in this model can account for the
experimental data on the heavy scalar diphoton resonance observed by the ATLAS
and CMS collaborations naturally. Simultaneously, this supersymmetric model can fit
the 125 GeV Higgs data determined by the earlier run I at the LHC.

It is proposed that besides the diphoton
channel the main decay portals are not to the SM particles, at least not at the tree
level, instead, it may decay into dark matter which is BSM particles. Moreover,
if the new physics scale is indeed at TeV, we have all reasons to expect observing
more resonances (charged and neutral) with some strange behaviors which cannot be understood
in the framework of the SM.

No doubt, the discovery of the diphoton excess at 750 GeV and confirmation of the
750 GeV resonance is a great breakthrough, but it is necessary to put more
efforts to investigate relevant physics. If eventually the 750 GeV is firmly identified as
a genuine particle which definitely is a BSM boson, a new world will be opened
in front of us, especially, the project to build up the SPPC of 50$\sim$100 TeV in China
should be more favorable and we are expecting the new spring of high energy physics to come soon.

\begin{acknowledgments}
\indent\indent
The work has been supported by the National Natural Science Foundation of China (NNSFC)
with Grants No. 11275036, No. 11535002,
the Open Project Program of State Key Laboratory of Theoretical Physics,
Institute of Theoretical Physics, Chinese Academy of Sciences, China (No. Y5KF131CJ1),
the Natural Science Foundation of Hebei province with Grants No. A2013201277, No. A2016201010, No. A2016201069,
the Natural Science Foundation of Hebei University with Grants No. 2011JQ05, No. 2012-242,
and Hebei Key Lab of Optic-Eletronic Information and Meterials.
\end{acknowledgments}

\appendix
\section{The radiative corrections to the mass squared matrix from exotic quark fields\label{app1}}
\indent\indent
The one-loop radiative corrections from
exotic quark fields are formulated as
\cite{1loopH,1loopH-1,1loopH-1-1,1loopH-2,1loopH-3,1loopH-4,1loopH-5,1loopH-6,1loopH-7}
\begin{eqnarray}
&&\Delta_{11}^{B}={3G_{_F}Y_{_{u_4}}^4\upsilon^4\over4\sqrt{2}\pi^2\sin^2\beta}\cdot
{\mu^2(A_{_{u_4}}-\mu\cot\beta)^2\over(m_{_{\tilde{t}_1^\prime}}^2-m_{_{\tilde{t}_2^\prime}}^2)^2}
g(m_{_{\tilde{t}_1^\prime}},m_{_{\tilde{t}_2^\prime}})
\nonumber\\
&&\hspace{1.2cm}
+{3G_{_F}Y_{_{u_5}}^4\upsilon^4\over4\sqrt{2}\pi^2\cos^2\beta}\Big\{\ln{m_{_{\tilde{t}_3^\prime}}m_{_{\tilde{t}_4^\prime}}
\over m_{_{t_5}}^2}+{A_{_{u_5}}(A_{_{u_5}}-\mu\tan\beta)\over m_{_{\tilde{t}_3^\prime}}^2-m_{_{\tilde{t}_4^\prime}}^2}
\ln{m_{_{\tilde{t}_3^\prime}}^2\over m_{_{\tilde{t}_4^\prime}}^2}
\nonumber\\
&&\hspace{1.2cm}
+{A_{_{u_5}}^2(A_{_{u_5}}-\mu\tan\beta)^2\over(m_{_{\tilde{t}_3^\prime}}^2-m_{_{\tilde{t}_4^\prime}}^2)^2}
g(m_{_{\tilde{t}_3^\prime}},m_{_{\tilde{t}_4^\prime}})\Big\}
\nonumber\\
&&\hspace{1.2cm}
+{3G_{_F}Y_{_{d_4}}^4\upsilon^4\over4\sqrt{2}\pi^2\cos^2\beta}\Big\{\ln{m_{_{\tilde{b}_1^\prime}}m_{_{\tilde{b}_2^\prime}}
\over m_{_{b_4}}^2}+{A_{_{d_4}}(A_{_{d_4}}-\mu\tan\beta)\over m_{_{\tilde{b}_1^\prime}}^2-m_{_{\tilde{b}_2^\prime}}^2}
\ln{m_{_{\tilde{b}_1^\prime}}^2\over m_{_{\tilde{b}_2^\prime}}^2}
\nonumber\\
&&\hspace{1.2cm}
+{A_{_{d_4}}^2(A_{_{d_4}}-\mu\tan\beta)^2\over(m_{_{\tilde{b}_1^\prime}}^2-m_{_{\tilde{b}_2^\prime}}^2)^2}
g(m_{_{\tilde{b}_1^\prime}},m_{_{\tilde{b}_2^\prime}})\Big\}
\nonumber\\
&&\hspace{1.2cm}
+{3G_{_F}Y_{_{d_5}}^4\upsilon^4\over4\sqrt{2}\pi^2\sin^2\beta}\cdot
{\mu^2(A_{_{d_5}}-\mu\cot\beta)^2\over(m_{_{\tilde{b}_3^\prime}}^2-m_{_{\tilde{b}_4^\prime}}^2)^2}
g(m_{_{\tilde{b}_3^\prime}},m_{_{\tilde{b}_4^\prime}})
\;,\\
&&\Delta_{12}^{B}={3G_{_F}Y_{_{u_4}}^4\upsilon^4\over8\sqrt{2}\pi^2\sin^2\beta}\cdot
{\mu(-A_{_{u_4}}+\mu\cot\beta)\over m_{_{\tilde{t}_1^\prime}}^2-m_{_{\tilde{t}_2^\prime}}^2}
\Big\{\ln{m_{_{\tilde{t}_1^\prime}}\over m_{_{\tilde{t}_2^\prime}}}+{A_{_{u_4}}(A_{_{u_4}}-\mu\cot\beta)
\over m_{_{\tilde{t}_1^\prime}}^2-m_{_{\tilde{t}_2^\prime}}^2}g(m_{_{\tilde{t}_1^\prime}},m_{_{\tilde{t}_2^\prime}})\Big\}
\nonumber\\
&&\hspace{1.2cm}
+{3G_{_F}Y_{_{u_5}}^4\upsilon^4\over8\sqrt{2}\pi^2\cos^2\beta}\cdot
{\mu(-A_{_{u_5}}+\mu\tan\beta)\over m_{_{\tilde{t}_3^\prime}}^2-m_{_{\tilde{t}_4^\prime}}^2}
\Big\{\ln{m_{_{\tilde{t}_3^\prime}}\over m_{_{\tilde{t}_4^\prime}}}+{A_{_{u_5}}(A_{_{u_5}}-\mu\tan\beta)
\over m_{_{\tilde{t}_3^\prime}}^2-m_{_{\tilde{t}_4^\prime}}^2}g(m_{_{\tilde{t}_3^\prime}},m_{_{\tilde{t}_4^\prime}})\Big\}
\nonumber\\
&&\hspace{1.2cm}
+{3G_{_F}Y_{_{d_4}}^4\upsilon^4\over8\sqrt{2}\pi^2\cos^2\beta}\cdot
{\mu(-A_{_{d_4}}+\mu\tan\beta)\over m_{_{\tilde{d}_1^\prime}}^2-m_{_{\tilde{d}_2^\prime}}^2}
\Big\{\ln{m_{_{\tilde{d}_1^\prime}}\over m_{_{\tilde{d}_2^\prime}}}+{A_{_{d_4}}(A_{_{d_4}}-\mu\tan\beta)
\over m_{_{\tilde{d}_1^\prime}}^2-m_{_{\tilde{d}_2^\prime}}^2}g(m_{_{\tilde{d}_1^\prime}},m_{_{\tilde{d}_2^\prime}})\Big\}
\nonumber\\
&&\hspace{1.2cm}
+{3G_{_F}Y_{_{d_5}}^4\upsilon^4\over8\sqrt{2}\pi^2\sin^2\beta}\cdot
{\mu(-A_{_{d_5}}+\mu\cot\beta)\over m_{_{\tilde{b}_3^\prime}}^2-m_{_{\tilde{b}_4^\prime}}^2}
\Big\{\ln{m_{_{\tilde{b}_3^\prime}}\over m_{_{\tilde{b}_4^\prime}}}+{A_{_{d_5}}(A_{_{d_5}}-\mu\cot\beta)
\over m_{_{\tilde{b}_3^\prime}}^2-m_{_{\tilde{b}_4^\prime}}^2}g(m_{_{\tilde{b}_3^\prime}},m_{_{\tilde{b}_4^\prime}})\Big\}
\;,\nonumber\\
&&\\
&&\Delta_{22}^{B}={3G_{_F}Y_{_{u_4}}^4\upsilon^4\over4\sqrt{2}\pi^2\sin^2\beta}\Big\{\ln{m_{_{\tilde{t}_1^\prime}}m_{_{\tilde{t}_2^\prime}}
\over m_{_{t_4}}^2}+{A_{_{u_4}}(A_{_{u_4}}-\mu\cot\beta)\over m_{_{\tilde{t}_1^\prime}}^2-m_{_{\tilde{t}_2^\prime}}^2}
\ln{m_{_{\tilde{t}_1^\prime}}^2\over m_{_{\tilde{t}_2^\prime}}^2}
\nonumber\\
&&\hspace{1.2cm}
+{A_{_{u_4}}^2(A_{_{u_4}}-\mu\cot\beta)^2\over(m_{_{\tilde{t}_1^\prime}}^2-m_{_{\tilde{t}_2^\prime}}^2)^2}
g(m_{_{\tilde{t}_1^\prime}},m_{_{\tilde{t}_2^\prime}})\Big\}
\nonumber\\
&&\hspace{1.2cm}
+{3G_{_F}Y_{_{u_5}}^4\upsilon^4\over4\sqrt{2}\pi^2\cos^2\beta}\cdot
{\mu^2(A_{_{u_5}}-\mu\tan\beta)^2\over(m_{_{\tilde{t}_3^\prime}}^2-m_{_{\tilde{t}_4^\prime}}^2)^2}
g(m_{_{\tilde{t}_3^\prime}},m_{_{\tilde{t}_4^\prime}})
\nonumber\\
&&\hspace{1.2cm}
+{3G_{_F}Y_{_{d_4}}^4\upsilon^4\over4\sqrt{2}\pi^2\cos^2\beta}\cdot
{\mu^2(A_{_{d_4}}-\mu\tan\beta)^2\over(m_{_{\tilde{b}_1^\prime}}^2-m_{_{\tilde{b}_2^\prime}}^2)^2}
g(m_{_{\tilde{b}_1^\prime}},m_{_{\tilde{b}_2^\prime}})
\nonumber\\
&&\hspace{1.2cm}
+{3G_{_F}Y_{_{d_5}}^4\upsilon^4\over4\sqrt{2}\pi^2\sin^2\beta}\Big\{\ln{m_{_{\tilde{b}_3^\prime}}m_{_{\tilde{b}_4^\prime}}
\over m_{_{b_5}}^2}+{A_{_{d_5}}(A_{_{d_5}}-\mu\cot\beta)\over m_{_{\tilde{b}_3^\prime}}^2-m_{_{\tilde{b}_4^\prime}}^2}
\ln{m_{_{\tilde{b}_3^\prime}}^2\over m_{_{\tilde{b}_4^\prime}}^2}
\nonumber\\
&&\hspace{1.2cm}
+{A_{_{d_5}}^2(A_{_{d_5}}-\mu\cot\beta)^2\over(m_{_{\tilde{b}_3^\prime}}^2-m_{_{\tilde{b}_4^\prime}}^2)^2}
g(m_{_{\tilde{b}_3^\prime}},m_{_{\tilde{b}_4^\prime}})\Big\}\;,
\label{M-CPE3}
\end{eqnarray}
here $\upsilon=\sqrt{\upsilon_{_u}^2+\upsilon_{_d}^2}\simeq246\;{\rm GeV}$ and
\begin{eqnarray}
&&g(x,y)=1-{x^2+y^2\over x^2-y^2}\ln{x\over y}\;.
\label{M-CPE4}
\end{eqnarray}
To derive the results presented above, we adopt the appropriate assumptions $|\lambda_{_Q}\upsilon_{_B}|,\;
|\lambda_{_U}\overline{\upsilon}_{_B}|,\;|\lambda_{_D}\overline{\upsilon}_{_B}|\gg |Y_{_{u_4}}\upsilon|,\;
|Y_{_{u_5}}\upsilon|,\;|Y_{_{d_4}}\upsilon|,\;|Y_{_{d_5}}\upsilon|$ in our calculation.

\section{The couplings between heavy Higgs and exotic quarks/squarks\label{PhiquarkCoupling}}
\indent\indent
\begin{eqnarray}
&&g_{_{H^0t_{(i+3)}t_{(i+3)}}}=-{\sqrt{2}m_{_{\rm W}}s_{_{\rm W}}\over em_{_{t_{(i+3)}}}}
\Big[Y_{_{u_4}}(W_{_t}^\dagger)_{_{i2}}(U_{_t})_{_{1i}}\sin\alpha
-Y_{_{u_5}}(W_{_t}^\dagger)_{_{i1}}(U_{_t})_{_{2i}}\cos\alpha\Big]
\;,\nonumber\\
&&g_{_{H^0b_{(i+3)}b_{(i+3)}}}={\sqrt{2}m_{_{\rm W}}s_{_{\rm W}}\over em_{_{b_{(i+3)}}}}
\Big[Y_{_{d_4}}(W_{_b}^\dagger)_{_{i2}}(U_{_b})_{_{1i}}\cos\alpha
+Y_{_{d_5}}(W_{_b}^\dagger)_{_{i1}}(U_{_b})_{_{2i}}\sin\alpha\Big]
\;,\nonumber\\
&&g_{_{H^0\tilde{\cal U}_i\tilde{\cal U}_i}}=-{s_{_{\rm W}}c_{_{\rm W}} \over e m_{_{\rm Z}}}
\Big[\xi_{_{uii}}^S\sin\alpha+\xi_{_{dii}}^S\cos\alpha\Big]\;,\;\;(i=1,\;2,\;3,\;4)
\;,\nonumber\\
&&g_{_{H^0\tilde{\cal D}_i\tilde{\cal D}_i}}=-{s_{_{\rm W}}c_{_{\rm W}} \over e m_{_{\rm Z}}}
\Big[\eta_{_{uii}}^S\sin\alpha+\eta_{_{dii}}^S\cos\alpha\Big]\;,\;\;(i=1,\;2,\;3,\;4)\;.
\label{g-coupling1}
\end{eqnarray}
\begin{eqnarray}
&&g_{_{h_{_B}^0t_{(i+3)}t_{(i+3)}}}={\sqrt{2}m_{_{\rm W}}s_{_{\rm W}}\over em_{_{t_{(i+3)}}}}
\Big[\lambda_{_U}(W_{_t}^\dagger)_{_{i2}}(U_{_t})_{_{2i}}\cos\alpha_{_B}
-\lambda_{_Q}(W_{_t}^\dagger)_{_{i1}}(U_{_t})_{_{1i}}\sin\alpha_{_B}\Big]
\;,\nonumber\\
&&g_{_{h_{_B}^0b_{(i+3)}b_{(i+3)}}}={\sqrt{2}m_{_{\rm W}}s_{_{\rm W}}\over em_{_{b_{(i+3)}}}}
\Big[\lambda_{_D}(W_{_b}^\dagger)_{_{i2}}(U_{_b})_{_{2i}}\cos\alpha_{_B}
+\lambda_{_Q}(W_{_b}^\dagger)_{_{i1}}(U_{_b})_{_{1i}}\sin\alpha_{_B}\Big]
\;,\nonumber\\
&&g_{_{h_{_B}^0\tilde{\cal U}_i\tilde{\cal U}_i}}=-{s_{_{\rm W}}c_{_{\rm W}} \over e m_{_{\rm Z}}}
\Big[\varsigma_{_{uii}}^S\cos\alpha_{_B}-\varsigma_{_{dii}}^S\sin\alpha_{_B}\Big]\;,\;\;(i=1,\;2,\;3,\;4)
\;,\nonumber\\
&&g_{_{h_{_B}^0\tilde{\cal D}_i\tilde{\cal D}_i}}=-{s_{_{\rm W}}c_{_{\rm W}} \over e m_{_{\rm Z}}}
\Big[\zeta_{_{uii}}^S\cos\alpha_{_B}-\zeta_{_{dii}}^S\sin\alpha_{_B}\Big]\;,\;\;(i=1,\;2,\;3,\;4)\;.
\label{g-coupling2}
\end{eqnarray}
\begin{eqnarray}
&&g_{_{H_{_B}^0t_{(i+3)}t_{(i+3)}}}={\sqrt{2}m_{_{\rm W}}s_{_{\rm W}}\over em_{_{t_{(i+3)}}}}
\Big[\lambda_{_U}(W_{_t}^\dagger)_{_{i2}}(U_{_t})_{_{2i}}\sin\alpha_{_B}
+\lambda_{_Q}(W_{_t}^\dagger)_{_{i1}}(U_{_t})_{_{1i}}\cos\alpha_{_B}\Big]
\;,\nonumber\\
&&g_{_{H_{_B}^0b_{(i+3)}b_{(i+3)}}}={\sqrt{2}m_{_{\rm W}}s_{_{\rm W}}\over em_{_{b_{(i+3)}}}}
\Big[\lambda_{_D}(W_{_b}^\dagger)_{_{i2}}(U_{_b})_{_{2i}}\sin\alpha_{_B}
+\lambda_{_Q}(W_{_b}^\dagger)_{_{i1}}(U_{_b})_{_{1i}}\cos\alpha_{_B}\Big]
\;,\nonumber\\
&&g_{_{H_{_B}^0\tilde{\cal U}_i\tilde{\cal U}_i}}=-{s_{_{\rm W}}c_{_{\rm W}} \over e m_{_{\rm Z}}}
\Big[\varsigma_{_{uii}}^S\sin\alpha_{_B}+\varsigma_{_{dii}}^S\cos\alpha_{_B}\Big]\;,\;\;(i=1,\;2,\;3,\;4)
\;,\nonumber\\
&&g_{_{H_{_B}^0\tilde{\cal D}_i\tilde{\cal D}_i}}=-{s_{_{\rm W}}c_{_{\rm W}} \over e m_{_{\rm Z}}}
\Big[\zeta_{_{uii}}^S\sin\alpha_{_B}+\zeta_{_{dii}}^S\cos\alpha_{_B}\Big]\;,\;\;(i=1,\;2,\;3,\;4)\;.
\label{g-coupling3}
\end{eqnarray}
\begin{eqnarray}
&&g_{_{A^0t_{(i+3)}t_{(i+3)}}}=-{\sqrt{2}m_{_{\rm W}}s_{_{\rm W}}\over em_{_{t_{(i+3)}}}}
\Big[Y_{_{u_4}}(W_{_t}^\dagger)_{_{i2}}(U_{_t})_{_{1i}}\cos\beta
+Y_{_{u_5}}(W_{_t}^\dagger)_{_{i1}}(U_{_t})_{_{2i}}\sin\beta\Big]
\;,\nonumber\\
&&g_{_{A^0b_{(i+3)}b_{(i+3)}}}={\sqrt{2}m_{_{\rm W}}s_{_{\rm W}}\over em_{_{b_{(i+3)}}}}
\Big[Y_{_{d_4}}(W_{_b}^\dagger)_{_{i2}}(U_{_b})_{_{1i}}\sin\beta
-Y_{_{d_5}}(W_{_b}^\dagger)_{_{i1}}(U_{_b})_{_{2i}}\cos\beta\Big]
\;.
\label{g-coupling4}
\end{eqnarray}
\begin{eqnarray}
&&g_{_{A_{_B}^0t_{(i+3)}t_{(i+3)}}}=-{\sqrt{2}m_{_{\rm W}}s_{_{\rm W}}\over em_{_{t_{(i+3)}}}}
\Big[\lambda_{_U}(W_{_t}^\dagger)_{_{i2}}(U_{_t})_{_{2i}}\cos\beta_{_B}
-\lambda_{_Q}(W_{_t}^\dagger)_{_{i1}}(U_{_t})_{_{1i}}\sin\beta_{_B}\Big]
\;,\nonumber\\
&&g_{_{A_{_B}^0b_{(i+3)}b_{(i+3)}}}={\sqrt{2}m_{_{\rm W}}s_{_{\rm W}}\over em_{_{b_{(i+3)}}}}
\Big[\lambda_{_D}(W_{_b}^\dagger)_{_{i2}}(U_{_b})_{_{2i}}\sin\beta_{_B}
+\lambda_{_Q}(W_{_b}^\dagger)_{_{i1}}(U_{_b})_{_{1i}}\cos\beta_{_B}\Big]\;.
\label{g-coupling5}
\end{eqnarray}
Here, we adopt the abbreviation $s_{_{\rm W}}\equiv \sin\theta_{_{\rm W}}$ with $\theta_{_{\rm W}}$
being the Weinberg angle. Furthermore, $e$ is the electromagnetic coupling constant, and
the concrete expressions of $\xi_{_{uii}}^S,\;\xi_{_{dii}}^S,\;\eta_{_{uii}}^S,\;\eta_{_{dii}}^S$ can be
found in Ref.~\cite{Feng}.

\section{The loop induced couplings\label{LoopInduceCoupling}}
\indent\indent
The loop induced couplings $g_{_{\Phi ZZ}},\;g_{_{\Phi WW}}\;(\Phi=h_{_B}^0,\;H_{_B}^0,\;A_{_B}^0)$
are written as

\begin{eqnarray}
&&g_{_{h_{_B}^0WW}}={e Y_{_{u_4}}\lambda_{_Q}\upsilon_{_u}\over4(4\pi)^2s_{_{\rm W}}m_{_{\rm W}}}
\sin\alpha_{_B}\Big({11\over6}+\ln{\lambda_{_Q}^2\upsilon_{_B}^2\over\Lambda_{_{NP}}^2}\Big)
\nonumber\\
&&\hspace{1.5cm}
+{e Y_{_{u_5}}\lambda_{_U}\upsilon_{_d}\over4(4\pi)^2s_{_{\rm W}} m_{_{\rm W}}}
\cos\alpha_{_B}\Big({11\over6}+\ln{\lambda_{_U}^2\bar{\upsilon}_{_B}^2\over\Lambda_{_{NP}}^2}\Big)
\nonumber\\
&&\hspace{1.5cm}
+{e Y_{_{d_4}}\lambda_{_Q}\upsilon_{_d}\over4(4\pi)^2s_{_{\rm W}} m_{_{\rm W}}}
\sin\alpha_{_B}\Big({11\over6}+\ln{\lambda_{_Q}^2\upsilon_{_B}^2\over\Lambda_{_{NP}}^2}\Big)
\nonumber\\
&&\hspace{1.5cm}
+{e Y_{_{d_5}}\lambda_{_D}\upsilon_{_u}\over4(4\pi)^2s_{_{\rm W}} m_{_{\rm W}}}
\cos\alpha_{_B}\Big({11\over6}+\ln{\lambda_{_D}^2\bar{\upsilon}_{_B}^2\over\Lambda_{_{NP}}^2}\Big)
\nonumber\\
&&\hspace{1.5cm}
-{B_{_4}e g_{_B}^2\over4(4\pi)^2s_{_{\rm W}} m_{_{\rm W}}}(\upsilon_{_B}\sin\alpha_{_B}
+\bar{\upsilon}_{_B}\cos\alpha_{_B})
\Big(2+\ln{m_{_{{\tilde Q}_4}}^2\over\Lambda_{_{NP}}^2}\Big)
\nonumber\\
&&\hspace{1.5cm}
+{(1+B_{_4})e g_{_B}^2\over4(4\pi)^2s_{_{\rm W}} m_{_{\rm W}}}(\upsilon_{_B}\sin\alpha_{_B}
+\bar{\upsilon}_{_B}\cos\alpha_{_B})\Big(2+\ln{m_{_{{\tilde Q}_5}}^2\over\Lambda_{_{NP}}^2}\Big)
\;,\\
&&g_{_{h_{_B}^0ZZ}}={e Y_{_{u_4}}\lambda_{_Q}\upsilon_{_u}\over36(4\pi)^2s_{_{\rm W}} c_{_{\rm W}} m_{_{\rm Z}}}
(3-4s_{_{\rm W}}^2)^2\sin\alpha_{_B}\Big({11\over6}+\ln{\lambda_{_Q}^2\upsilon_{_B}^2\over\Lambda_{_{NP}}^2}\Big)
\nonumber\\
&&\hspace{1.5cm}
+{e Y_{_{u_5}}\lambda_{_U}\upsilon_{_d}\over36(4\pi)^2s_{_{\rm W}} c_{_{\rm W}} m_{_{\rm Z}}}
(3-4s_{_{\rm W}}^2)^2\cos\alpha_{_B}\Big({11\over6}+\ln{\lambda_{_U}^2\bar{\upsilon}_{_B}^2\over\Lambda_{_{NP}}^2}\Big)
\nonumber\\
&&\hspace{1.5cm}
+{e Y_{_{d_4}}\lambda_{_Q}\upsilon_{_d}\over36(4\pi)^2s_{_{\rm W}} c_{_{\rm W}} m_{_{\rm Z}}}
(3-2s_{_{\rm W}}^2)^2\sin\alpha_{_B}\Big({11\over6}+\ln{\lambda_{_Q}^2\upsilon_{_B}^2\over\Lambda_{_{NP}}^2}\Big)
\nonumber\\
&&\hspace{1.5cm}
+{e Y_{_{d_5}}\lambda_{_D}\upsilon_{_u}\over36(4\pi)^2s_{_{\rm W}} c_{_{\rm W}} m_{_{\rm Z}}}
(3-2s_{_{\rm W}}^2)^2\cos\alpha_{_B}\Big({11\over6}+\ln{\lambda_{_D}^2\bar{\upsilon}_{_B}^2\over\Lambda_{_{NP}}^2}\Big)
\nonumber\\
&&\hspace{1.5cm}
-{B_{_4}e g_{_B}^2\over4(4\pi)^2s_{_{\rm W}} c_{_{\rm W}} m_{_{\rm Z}}}(2-4s_{_{\rm W}}^2+{20\over9}s_{_{\rm W}}^4)
(\upsilon_{_B}\sin\alpha_{_B}
\nonumber\\
&&\hspace{1.5cm}
+\bar{\upsilon}_{_B}\cos\alpha_{_B})
\Big(2+\ln{m_{_{{\tilde Q}_4}}^2\over\Lambda_{_{NP}}^2}\Big)
\nonumber\\
&&\hspace{1.5cm}
+{(1+B_{_4})e g_{_B}^2\over4(4\pi)^2s_{_{\rm W}} c_{_{\rm W}} m_{_{\rm Z}}}(2-4s_{_{\rm W}}^2
+{20\over9}s_{_{\rm W}}^4)(\upsilon_{_B}\sin\alpha_{_B}
\nonumber\\
&&\hspace{1.5cm}
+\bar{\upsilon}_{_B}\cos\alpha_{_B})\Big(2+\ln{m_{_{{\tilde Q}_5}}^2\over\Lambda_{_{NP}}^2}\Big)
\nonumber\\
&&\hspace{1.5cm}
+{4B_{_4}e g_{_B}^2s_{_{\rm W}}^3\over9(4\pi)^2c_{_{\rm W}} m_{_{\rm Z}}}
(\upsilon_{_B}\sin\alpha_{_B}+\bar{\upsilon}_{_B}\cos\alpha_{_B})
\Big(2+\ln{m_{_{{\tilde U}_4}}^2\over\Lambda_{_{NP}}^2}\Big)
\nonumber\\
&&\hspace{1.5cm}
+{B_{_4}e g_{_B}^2s_{_{\rm W}}^3\over9(4\pi)^2c_{_{\rm W}} m_{_{\rm Z}}}
(\upsilon_{_B}\sin\alpha_{_B}+\bar{\upsilon}_{_B}\cos\alpha_{_B})
\Big(2+\ln{m_{_{{\tilde D}_4}}^2\over\Lambda_{_{NP}}^2}\Big)
\nonumber\\
&&\hspace{1.5cm}
-{4(1+B_{_4})e g_{_B}^2s_{_{\rm W}}^3\over9(4\pi)^2c_{_{\rm W}} m_{_{\rm Z}}}
(\upsilon_{_B}\sin\alpha_{_B}+\bar{\upsilon}_{_B}\cos\alpha_{_B})
\Big(2+\ln{m_{_{{\tilde U}_5}}^2\over\Lambda_{_{NP}}^2}\Big)
\nonumber\\
&&\hspace{1.5cm}
-{(1+B_{_4})e g_{_B}^2s_{_{\rm W}}^3\over9(4\pi)^2c_{_{\rm W}} m_{_{\rm Z}}}
(\upsilon_{_B}\sin\alpha_{_B}+\bar{\upsilon}_{_B}\cos\alpha_{_B})
\Big(2+\ln{m_{_{{\tilde D}_5}}^2\over\Lambda_{_{NP}}^2}\Big)
\;,\\
&&g_{_{H_{_B}^0WW}}=g_{_{h_{_B}^0WW}}(\sin\alpha_{_B}\rightarrow-\cos\alpha_{_B},
\cos\alpha_{_B}\rightarrow\sin\alpha_{_B})
\;,\\
&&g_{_{H_{_B}^0ZZ}}=g_{_{h_{_B}^0ZZ}}(\sin\alpha_{_B}\rightarrow-\cos\alpha_{_B},
\cos\alpha_{_B}\rightarrow\sin\alpha_{_B})
\;,\\
&&g_{_{A_{_B}^0WW}}={ie Y_{_{u_4}}\lambda_{_Q}\upsilon_{_u}\over4(4\pi)^2s_{_{\rm W}} m_{_{\rm W}}}
\sin\beta_{_B}\Big({11\over6}+\ln{\lambda_{_Q}^2\upsilon_{_B}^2\over\Lambda_{_{NP}}^2}\Big)
\nonumber\\
&&\hspace{1.5cm}
+{ie Y_{_{u_5}}\lambda_{_U}\upsilon_{_d}\over4(4\pi)^2s_{_{\rm W}} m_{_{\rm W}}}
\cos\beta_{_B}\Big({11\over6}+\ln{\lambda_{_U}^2\bar{\upsilon}_{_B}^2\over\Lambda_{_{NP}}^2}\Big)
\nonumber\\
&&\hspace{1.5cm}
+{ie Y_{_{d_4}}\lambda_{_Q}\upsilon_{_d}\over4(4\pi)^2s_{_{\rm W}} m_{_{\rm W}}}
\sin\beta_{_B}\Big({11\over6}+\ln{\lambda_{_Q}^2\upsilon_{_B}^2\over\Lambda_{_{NP}}^2}\Big)
\nonumber\\
&&\hspace{1.5cm}
+{ie Y_{_{d_5}}\lambda_{_D}\upsilon_{_u}\over4(4\pi)^2s_{_{\rm W}} m_{_{\rm W}}}
\cos\beta_{_B}\Big({11\over6}+\ln{\lambda_{_D}^2\bar{\upsilon}_{_B}^2\over\Lambda_{_{NP}}^2}\Big)
\;,\\
&&g_{_{A_{_B}^0ZZ}}={ie Y_{_{u_4}}\lambda_{_Q}\upsilon_{_u}\over36(4\pi)^2s_{_{\rm W}} c_{_{\rm W}} m_{_{\rm Z}}}
(3-4s_{_{\rm W}}^2)^2\sin\beta_{_B}\Big({11\over6}+\ln{\lambda_{_Q}^2\upsilon_{_B}^2\over\Lambda_{_{NP}}^2}\Big)
\nonumber\\
&&\hspace{1.5cm}
+{ie Y_{_{u_5}}\lambda_{_U}\upsilon_{_d}\over36(4\pi)^2s_{_{\rm W}} c_{_{\rm W}} m_{_{\rm Z}}}
(3-4s_{_{\rm W}}^2)^2\cos\beta_{_B}\Big({11\over6}+\ln{\lambda_{_U}^2\bar{\upsilon}_{_B}^2\over\Lambda_{_{NP}}^2}\Big)
\nonumber\\
&&\hspace{1.5cm}
+{ie Y_{_{d_4}}\lambda_{_Q}\upsilon_{_d}\over36(4\pi)^2s_{_{\rm W}} c_{_{\rm W}} m_{_{\rm Z}}}
(3-2s_{_{\rm W}}^2)^2\sin\beta_{_B}\Big({11\over6}+\ln{\lambda_{_Q}^2\upsilon_{_B}^2\over\Lambda_{_{NP}}^2}\Big)
\nonumber\\
&&\hspace{1.5cm}
+{ie Y_{_{d_5}}\lambda_{_D}\upsilon_{_u}\over36(4\pi)^2s_{_{\rm W}} c_{_{\rm W}} m_{_{\rm Z}}}
(3-2s_{_{\rm W}}^2)^2\cos\beta_{_B}\Big({11\over6}+\ln{\lambda_{_D}^2\bar{\upsilon}_{_B}^2\over\Lambda_{_{NP}}^2}\Big).
\label{Phi-VV}
\end{eqnarray}

\end{document}